\documentclass[traditabstract]{aa} 
\usepackage{amsmath}
\usepackage{graphicx}
\usepackage{txfonts}

\usepackage{natbib}
\bibpunct{(}{)}{;}{a}{}{,}
\usepackage{float}
\usepackage{calc}
\usepackage{textcomp}
\usepackage{placeins}
\usepackage{color}
\usepackage{rotating} 
\usepackage{lscape}
\usepackage{url}
\usepackage[colorlinks=true,linkcolor=blue,citecolor=blue]{hyperref}
\usepackage{subfig}
\usepackage[all]{draftcopy}
\usepackage{longtable}
\usepackage{array}
\usepackage{threeparttable}
\usepackage{lscape}
\usepackage{enumerate}
\DeclareTextSymbol{\degre}{OT1}{23}
\newcounter{savedfootnote}

\renewcommand{\epsilon}{\varepsilon} 
\def \microns{{\,$\mu$m}}
\def \HI{{H{\sc i}}}

\begin{document}
\title{Investigating the delay between dust radiation and star-formation in
local and distant quenching galaxies}

\author{L.~Ciesla\inst{1},
        V.~Buat\inst{1,2},
        M.~Boquien\inst{3}, 
        A.~Boselli\inst{1},
        D.~Elbaz\inst{4},
        and G.~Aufort\inst{1}.
}

\institute{	
 Aix-Marseille  Universit\'e,  CNRS, LAM (Laboratoire d'Astrophysique de Marseille) UMR7326,  13388, Marseille, France
 \and
 Institut Universitaire de France (IUF)
 \and
 Centro de Astronomía (CITEVA), Universidad de Antofagasta, Avenida Angamos 601, Antofagasta, Chile
 \and
 AIM-Paris-Saclay, CEA/DSM/Irfu - CNRS - Université Paris Diderot, CEA-Saclay, Pt Courrier 131, 91191, Gif-sur-Yvette, France
}	

   \date{Received; accepted}

  \abstract
{
We investigate the timescale with which the infrared (IR) luminosity decreases after a complete and rapid quenching of star formation using observations of local and high-redshift galaxies.
From SED modelling, we derive the time since quenching of a sub-sample of 14 galaxies from the \textit{Herschel} Reference Survey suffering from ram-pressure stripping due to the environment of the Virgo cluster and of a sub-sample of 7 rapidly quenched COSMOS galaxies selected through a state-of-the-art statistical method already tested on the determination of galaxies' star formation history (SFH).
Three out of the 7 COSMOS galaxies have an optical spectra with no emission line, confirming their quenched nature.
Present physical properties of the combined sample (local plus high-redshift) are obtained as well as the past L$_{IR}$ of these galaxies, just before their quenching, from the long-term SFH properties.
This past L$_{IR}$ is shown to be consistent with the L$_{IR}$ of reference samples of normally star-forming galaxies with same stellar mass and redshift than each of our quenched galaxies.
We put constraints on the present to past IR luminosity ratio as a function of quenching time.
The two samples probe different dynamical ranges in terms of quenching age with the HRS galaxies exhibiting longer timescales (0.2-3\,Gyr) compared to the COSMOS one ($<100$\,Myr).
Assuming an exponential decrease of the L$_{IR}$ after quenching, the COSMOS quenched galaxies are consistent with short e-folding times less than a couple of hundreds of Myr while the properties of the HRS quenched galaxies are compatible with larger timescales of several hundreds of Myr.
For the HRS sample, this result is consistent with the known quenching mechanism that affected them, ram pressure stripping due to the environment.
For the COSMOS sample, different quenching processes are acting on short to intermediate timescales.
Processes such as galaxy mergers, disk instabilities or environmental effects can produce such strong star formation variability.
}


   \authorrunning{Ciesla et al.}
   \titlerunning{Delay between dust radiation and star formation in
local and distant quenching galaxies}

   \maketitle

\section{\label{intro}Introduction}

The infrared (IR) luminosity of a galaxy is a key parameter tightly linked to its star formation activity and directly converted to determine its star formation rate \citep[SFR, e.g.,][]{KennicuttEvans12}.
Although most of the time the true SFR of a galaxy agrees well with the SFR inferred from the IR luminosity for galaxies actively forming stars, there are some evolutionary phases when the two are discrepant \citep{Hayward14,Boquien14}.
This is the case when galaxies experience short-terms variations of their star formation history (SFH) such as rapid quenching.
As a consequence, when using the IR luminosity as an SFR indicator one could conclude that a particular galaxy is still forming stars, whereas star formation has actually recently been quenched \citep{Hayward14}.
This bias may have consequences when investigating the short-term SFH of galaxies in the context of the galaxies star-forming main sequence (MS) paradigm \citep{Elbaz07,Noeske07}, for instance.
The main consequence of this tight relation between the SFR and stellar mass of galaxies is that they are forming the bulk of their stars through steady-state processes rather than violent episodes of star formation.

Although the MS is found to hold up to $z$=4 \citep{Schreiber17} with little variation of its normalisation and shape with redshift \citep{Daddi07,Pannella09,Elbaz11,Rodighiero11,Speagle14,Whitaker14,Schreiber15,Gavazzi15,Tomczak16}, what is striking is that the scatter of the MS is found to be relatively constant at all masses and over cosmic time \citep{Guo13,Ilbert15,Schreiber15}.
Several studies have found a coherent variation of physical galaxy properties such as the gas fraction \citep{Magdis12}, Sersic index, and effective radius \citep{Wuyts11}, U-V color \citep[e.g.,][]{Salmi12} suggesting that the bulk of the scatter is related to physics and not measurement and model uncertainties. 
From an observational point of view, \cite{Elbaz18} showed that some massive compact galaxies exhibiting starburst galaxy properties (short depletion time and high IR surface density) can be found within the MS.
However they have different morphology and gas fraction compared to "true" starbursts (above the MS), indicating a different origin, possibly  being late-stage mergers of gas-rich galaxies.
This could be the sign of a possible recent movement of these galaxies from the starburst galaxies region back to the MS.
From a theoretical aspect, oscillations of the SFR resulting from a varying infall rate and compaction of star-formation have also been advocated to explain the MS scatter \citep[e.g.,][]{DekelBurkert14,Sargent14,Scoville16,Tacchella16}.
These variations must be small enough to keep the SFR of the galaxy within the MS scatter.
However, based on EAGLE simulations, \cite{Matthee19} showed that although individual galaxies can cross the MS multiple times during their evolution, the main tracks around which they oscillate is linked to their halo properties, i.e. galaxies above/below the MS at $z=0.1$ tend to have been above/below the MS for more than 1\,Gyr.
Using 150 zoom-in simulations of galaxies, \cite{Blank21} obtained consistent results.
Therefore, there is no consensus on the evolution of galaxies relative to the MS and accurate measurements of galaxies position, present and past, on the SFR-M$_*$ diagram are needed to shed light on their short-term evolution.

As a step to reach this goal, we aim, in this work, at putting some constraints on the timescales with which the IR luminosity decreases after a complete shutdown of star formation activity.
To do so, we recover the past IR luminosity of our studied galaxies, just before quenching, from broad band spectral energy distribution (SED) modelling as well as their past star-forming properties.
Here, the word ``quenching'' is used to reflect rapid quenching processes with timescales less than $\sim$1\,Gyr as opposed to slower mechanisms such as mass quenching for instance.

The paper is organised as follows:
In Sect.~\ref{local}, Sect.~\ref{sedmod}, and Sect.~\ref{seleclocal}, we describe the \textit{Herschel} Reference Survey \citep[HRS,][]{Boselli10a} local sample, the broad band SED modelling method, and the selection of rapidly quenched candidates among the HRS, respectively.
To extend our study in terms of luminosity and redshift, we select a complementary sample of rapidly quenched galaxies from the COSMOS survey in Sect.~\ref{cosmos}.
The evolution of the IR luminosity of both the local and high-redshift samples is presented in Sect.~\ref{evol} and discussed in Sect~\ref{discussion}.
Finally, our conclusions are provided in Sect.~\ref{conclusions}.
Throughout this paper, we assume an IMF of \cite{Salpeter55}.

\section{\label{local}The sample of local galaxies}
We use the HRS which is a combined volume- and flux- limited sample composed of local galaxies with a distance between 15 and 25\,Mpc.
The galaxies are selected according to their K-band magnitude, a reliable proxy for the total stellar mass \citep{Gavazzi96}. 
The sample contains 322\footnote{With respect to the original sample given in \cite{Boselli10a}, the galaxy HRS\,228 is removed from the complete sample because its updated redshift on NED indicates it as a background object.}  galaxies, among which 62 early-type and 260 late-type.
We refer the reader to \cite{Boselli10a} for additional information on the sample.
In this work we only consider the 260 late-type galaxies.

The HRS sample is well suited for this study, it is partly composed of sources that are part of the Virgo cluster. 
Entering the intra-cluster medium, these galaxies have their gas content stripped through ram pressure stripping, quantified through a deficit of \HI\ gas content, resulting in a quenching of their star formation activity in a time scale of a few hundreds of Myr to a couple of Gyr \citep[see for instance][]{Boselli16}.
Since we aim at studying the decrease of IR luminosity after the shutdown of star formation, these sources are good targets.
Furthermore, the wealth of ancillary data, both photometric and spectroscopic, available for the HRS galaxies is an asset and allows us to probe the SFH of the galaxies as well as their IR properties \citep[][and from the literature]{Bendo12a,Cortese12b,Ciesla12,Cortese14,Ciesla14,Boselli13,Boselli14a,Boselli15}.
The photometric bands used in this work are listed in Table~\ref{filt}.

A sample of high-redshift galaxies complementing the HRS local sample will be described later on in Sect.~\ref{cosmos}.

\begin{table}
	\centering
	\caption{HRS broad band set of filters.}
	\begin{tabular}{ l l c l }
	 \hline\hline
	Telescope/Camera & Filter Name & $\lambda_{mean}$(\microns) & Ref.\\ 
	\hline
	GALEX & FUV &  0.153 & a \\
				 & NUV & 0.231& a \\
				& U & 0.365 & b\\
				& B & 0.44 & b\\
	SDSS	& g & 0.475 & a\\
				& V & 0.55 & b\\
	SDSS	& r & 0.622 & a\\
	SDSS	& i	& 0.763 & a\\		 
	2MASS &  J & 1.25 & b \\
			 &  H & 1.65 & b \\
			 & Ks & 2.1 &  b \\
	\textit{Spitzer}& IRAC1& 3.6 & c\\
			& IRAC2 & 4.5  & c \\
			& IRAC4 & 8 & d \\
	WISE & 3	& 12  & d \\
			& 4 & 22 &  d\\
	\textit{Spitzer}		& MIPS1 & 24 & e\\
			& MIPS2 & 70  & e\\
	\textit{Herschel}& PACS green & 100 & f\\
			& PACS red & 160 & f\\
			& PSW & 250 & g\\
			& PMW & 350 & g\\
			& PLW & 500 & g\\
	\hline
	\label{filt}
	\end{tabular}
	\tablefoot{
	\tablefoottext{a}{\cite{Cortese12b}.}
	\tablefoottext{b}{Compilation from the literature, details are provided in \cite{Boselli10a}.}
	\tablefoottext{c}{S$^4$G: \cite{Querejeta15}.}
	\tablefoottext{d}{\cite{Ciesla14}.}
	\tablefoottext{e}{\cite{Bendo12a}.}
	\tablefoottext{f}{\cite{Cortese14}.}
	\tablefoottext{g}{\cite{Ciesla12}.}
	}
\end{table}

\section{\label{sedmod}Spectral energy distribution modeling}
\subsection{The CIGALE code}
We use the SED modelling and fitting code CIGALE\footnote{\url{http://cigale.lam.fr}} \citep{Boquien19}.
CIGALE models and fits the UV to sub-millimetre (submm) emission of galaxies assuming an energy balance between the emission absorbed by dust in UV-optical and re-emitted in IR.
It is a versatile code composed of modules modelling the star formation history of galaxies, the stellar emission, the dust emission, the Active Galactic Nucleus (AGN) contribution, as well as the radio emission of galaxies.
In CIGALE, the SFH can be handled through analytical functions or using simulated SFHs \citep{Boquien14,Ciesla15,Ciesla17}.

In a previous study, we investigated the use of simple analytical SFH forms to recover galaxy parameters \citep{Ciesla15}. 
A set of SFH from semi-analytical models with known associated properties (SFR, M$_*$, etc.) were used to test different SFH analytical forms to recover them (one or two exponentially declining SFH, and a delayed SFH used in this paper). 
The delayed SFH can recover the SFR and M$_*$ properties associated to these simulated SFHs even in bursty SFHs \citep[see for instance Fig. 7 of][]{Ciesla15}. 
More recently, in \cite{Ciesla17} we studied the ability of recovering the same properties in case of galaxies that recently experienced strong variations. 
For these sources, the addition of an extra flexibility in the recent SFH is needed to better recover the SFR. 
This was confirmed further by \cite{Aufort20} from a statistical approach. 
As in this paper we are only focusing on the latest SFH variation, we relied on these studies and use a delayed SFH associated with the recent flexibility as presented and described in \cite{Ciesla17} and \cite{Aufort20}.
The delayed-$\tau$ SFH is defined as:
\begin{equation}        
\mathrm{SFR}(t) \propto t \times exp(-t/\tau_{main})\\
\end{equation} 
\noindent where SFR is the star formation rate, $t$ the time, and $\tau_{main}$ is the e-folding time.
The flexible SFH is an extension of the delayed-$\tau$ model:
\begin{equation}        
\mathrm{SFR}(t) \propto
\begin{cases}
    t \times exp(-t/\tau_{main}), & \text{when}\ t\leq t_{flex} \\
    r_{\rm{SFR}} \times \mathrm{SFR}(t=t_{flex}), & \text{when}\ t>t_{flex} \\
\end{cases},
\end{equation}  
        
\noindent where $t_{flex}$ is the time at which the star formation is instantaneously affected, and $r_{\rm{SFR}}$ is the ratio between SFR$(t>t_{flex})$ and SFR$(t=t_{flex})$:
        
\begin{equation}
   r_{\rm{SFR}} = \frac{\mathrm{SFR}(t>t_{flex})}{\mathrm{SFR}(t_{flex})}.
\end{equation}

From this we can define $age_{flex}$ which is the $age$ of the galaxy minus $t_{flex}$.
As we will focus on quenched galaxies for the rest of the study, $age_{flex}$ will be named $age_{trunc}$, the age of the SFH truncation, for clarity.
In addition to the flexible delayed-$\tau$ SFH, the SEDs of our sample are fitted using the stellar population models of \cite{BruzualCharlot03}, the \cite{CharlotFall00} attenuation recipe, and the \cite{Dale14} dust emission models.

The goal of this paper is to recover past and present properties of a sample of galaxies.
However, the attenuation curve of the galaxies before and after the quenching is probably different, and the dust content and dust-star geometry should vary as well.
We choose to use the \cite{CharlotFall00} attenuation law where a different attenuation is assumed for young ($<10^7$ years) and old ($>10^7$ years) stars. Light from both young and old stars is attenuated by the ISM but the young stars emission is also affected by the dust in the birth clouds (BC). 
Both attenuation, ISM and BC, are modelled by power laws with n$_{ISM}$ and n$_{BC}$. 
Furthermore, there is a $\mu$ parameter, defined as $\mu = A_V^{ISM}/(A_V^{ISM}+A_V^{BC})$ that can be used to handle the attenuation of old and young populations and thus change the effective attenuation law \citep[e.g.][]{Battisti20}. 
The difference in dust geometry is thus handled through this parameter. 
Except for the $A_V^{ISM}$ parameter which is free in our SED fitting procedure, n$_{ISM}$, n$_{BC}$ and $\mu$ are fixed to -0.7, -0.7, and 0.3 respectively. 
We tested several runs of SED fitting varying these three parameters: (n$_{ISM}$, n$_{BC}$, $\mu$) combinations of (-0,7; -0.7; 0.3), (-1; -1; 0.3), and (-0.7,-1.3; 0.3), and the same combinations with $\mu$ variable (0.2, 0.3, 0.4). 
For each test and for the quenched candidates, we compared the quality of the new fit to the one proposed in our paper using the Bayesian Information Criterion \citep[see for instance][]{Ciesla18,Aufort20}. 
The fits with different combinations did not result into a better quality fit compared to the one adopted here. 
Varying $\mu$ did not result into a better fit of the data despite the additional degree of freedom. 
However, we note that the \cite{CharlotFall00} attenuation law intrinsically takes into account some variations in the attenuation curves with time due to stellar populations ages.

Input parameters used for each modules are provided in Table~\ref{inputparam}.

\begin{table}
   \centering
   \caption{Input parameters used in the SED fitting procedures with CIGALE.}
   \begin{tabular}{l  l }
   \hline\hline
   Parameter & Value \\ 
   \hline\hline
   \multicolumn{2}{c}{Flexible delayed-$\tau$ SFH}\\  
   \hline
   $age$ (Gyr) &  11, 11.5, 12, 12.5, 13        \\
   $\tau_{main}$  (Gyr) & $[2;17]$, 5 values linearly sampled\\
   $age_{trunc}$ (Myr) & $[10;3000]$, 30 values log-sampled\\
   $r_{\rm{SFR}}$  & $[10^{-3};10^{2}]$, 20 values log-sampled     \\ 
   \hline\hline
   \multicolumn{2}{c}{Dust attenuation: \cite{CharlotFall00}}\\  
   \hline
   A$_V$ ISM      &  $[0.2;2.2]$, 15 values linearly sampled\\
   $\mu$      &  0.3\\
   $n_{ISM}$ & -0.7 \\
   $n_{\mathrm{Birth Cloud}}$ & -0.7 \\
   \hline\hline
   \multicolumn{2}{c}{Dust emission: \cite{Dale14}}\\  
   \hline
   $\alpha$      &  1.5, 2.0, 2.5\\
   \hline
   \hline
   \label{inputparam}
   \end{tabular}
\end{table}

\subsection{The SED fitting procedure}
To reach the goal of this study, that is to put constraints on the decrease of IR luminosity after rapid quenching of star formation, we perform the SED fitting in two steps.
First, we model the whole SED from UV to submm of the 260 star-forming galaxies of the HRS.
This run allows us to estimate the age at which a recent variation of the SFH, if needed, occurred.
We then reconstruct the SFH of the galaxies to estimate the IR luminosity of the galaxies right before the recent variation of SFH, as we will explain in Sect.~\ref{evol}.
In a second step, we fit only the IR part of the SED with data from MIPS 24\,$\mu$m to \textit{Herschel}/SPIRE 500\,$\mu$m to obtain a measurement of the current IR luminosity independently from the UV-NIR SED of the galaxy.

\subsection{\label{cons}Constraints on the parameters}

\begin{figure*}[!h] 
  	\includegraphics[width=\textwidth]{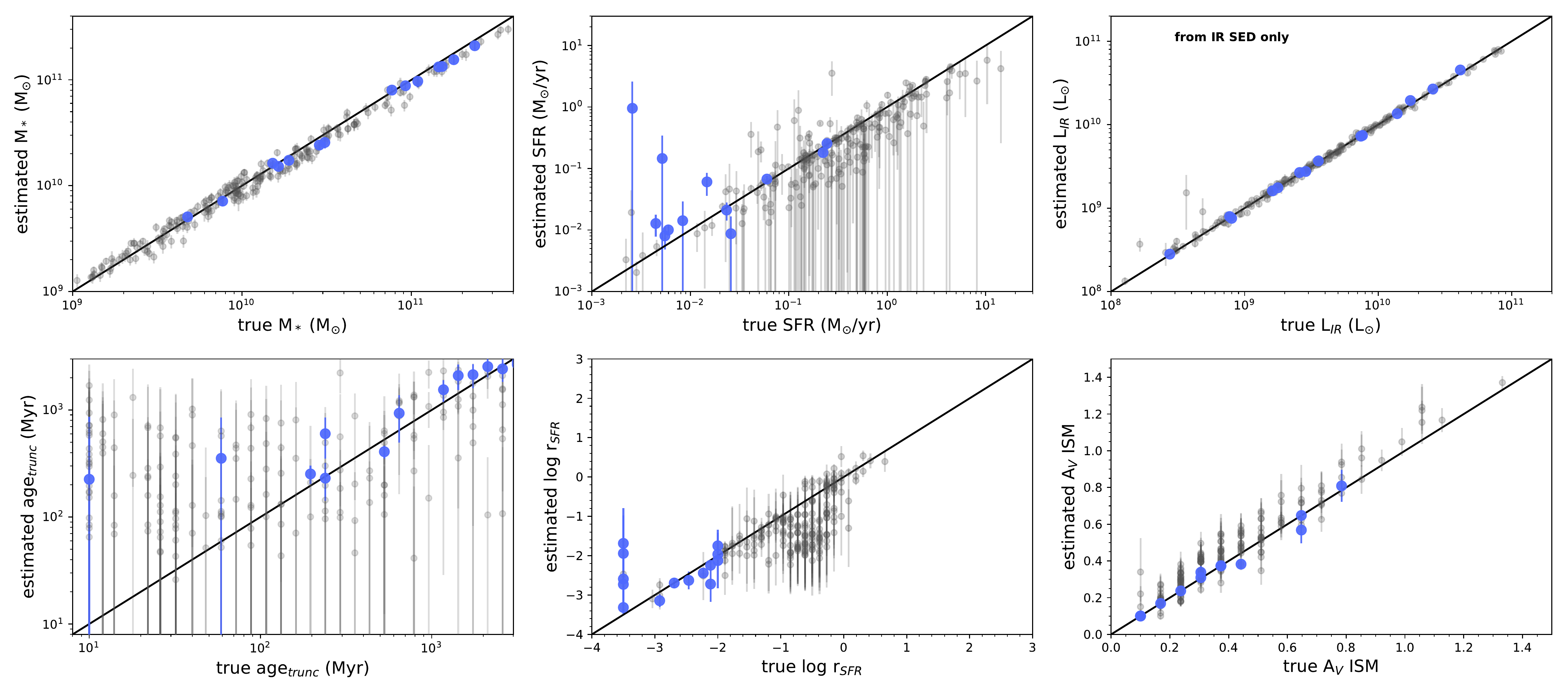}
  	\caption{\label{mocks_hrs} Results of the mock analysis. The input parameters used to build the mock catalogue are shown on the x-axis while the results of the fitting of the mock catalogues are shown on the y-axis. A good constraint on a given parameter is obtained when there is a one-to-one relationship, which is indicated by the black solid lines. Grey dots are all HRS sources sample while blue circles are the galaxies selected as rapidly quenched.}
\end{figure*}

Before analysing the results from SED fitting, one must know if the measured parameters are indeed constrained by the data.
To do so, we perform a mock analysis with CIGALE, a functionality available in the code.
The procedure is explained in, for instance, \cite{Giovannoli11} and \cite{Boquien19}, we summarise here the main steps.
A first run is made with CIGALE from which we obtain the best model for each galaxy, as well as the corresponding physical properties (stellar mass, SFR, age, etc...).
The best SED model of each galaxy, for which we know all parameters, is integrated into the same set of filters as our observed sample.
These mock flux densities are then perturbed by adding a noise randomly selected in a Gaussian distribution with a $\sigma$ corresponding to the error of the original flux density.
CIGALE, in the same configuration, is then run on this new mock catalogue for which each parameter is known.
The mock analysis then consists in comparing the results of the Bayesian-like analysis provided by CIGALE on this mock catalogue to the input parameters used to build it. 
If there is a one-to-one relationship between the input and output values of a parameter then it is perfectly constrained by the data in hand.
We use this test to check the robustness of the output SED fitting parameters, our ability to constrain them, and the possible degeneracies that can arise.

The results of the mock analysis performed on our HRS sample is shown in Fig.~\ref{mocks_hrs} (grey symbols).
The stellar mass is well constrained which is expected since there is a good coverage of the NIR wavelength range.
Overall SFRs are well recovered which is also expected since our sample benefit from a good UV to IR coverage \citep{Buat14}, with the exception of a couple of sources.
For the same reason, the V band attenuation parameter, A$_V$, is also well estimated.

Regarding the parameters linked to the SFH, we find that $\tau_{main}$ is not well constrained showing a dispersed relation, as discussed in previous studies \citep[e.g.,][]{Buat14,Ciesla16}.
For the $r_{\mathrm{SFR}}$ parameter, the relation between the input value and the one recovered by CIGALE does not follow exactly the one-to-one relationship.
Indeed, at lowest and highest input values, the relation is flat.
Lowest input values are overestimated, and highest values are underestimated.
This could be due to a well-known effect of PDF analysis:
the value estimated from the Bayesian-like analysis comes from the probability distribution function (PDF) of the parameter.
The final value is the mean of the PDF while the error is its standard deviation.
However, for the extreme values (lowest and highest) this PDF is truncated and therefore the mean value is skewed toward higher parameter value and lower parameter value for the lowest and highest input parameter values, respectively \citep[e.g.][]{Noll09,Buat12,Ciesla15}.
Furthermore values between 0.1 and 1 tend to be underestimated.
Another possibility can be that for very low values of $r_{\mathrm{SFR}}$ the spectrum of the galaxy does not show a lot of variations and becomes insensitive to the parameter, hence the PDF becomes flat.
We will discuss in the following how we choose our selection criteria to minimise biases linked to the difficulty in constraining $r_{\mathrm{SFR}}$.
We note that for most of the galaxies, the $r_{\mathrm{SFR}}$ seems to be underestimated by a factor of about 10.
We will take this into account while defining our selection criteria in the following.

Finally, the $age_{\mathrm{trunc}}$ parameter shows a dispersed distribution for the whole sample.
This is not surprising as the majority of the HRS galaxies are normal star-forming galaxies with a quasi constant SFH over the last several Gyr.
Therefore for these galaxies the truncated SFH is not well-suited, hence the difficulty to constrain $age_{\mathrm{trunc}}$.
The same problem can be at the origin of difficulty to constrain $r_{\mathrm{SFR}}$ as well for normal galaxies.
However, we will show in the following that $age_{\mathrm{trunc}}$ is well recovered for quenched galaxies.

The IR luminosity, obtained from the fit of the IR SED only, is very well constrained and recover by the SED fitting which is not surprising given the good IR coverage.

\section{\label{seleclocal}A fiducial sample of local quenched galaxies}

\subsection{Selection of HRS quenched galaxies}

We choose to apply a cut in $r_{\mathrm{SFR}}$ to select galaxies that are close to be totally quenched.
Based on the mock analysis described in the previous section, by selecting galaxies with $r_{SFR}\leq0.01$, that is sources for which the SFR after quenching is lower by a factor larger than 100, we ensure a conservative sample.
Indeed, as shown in Fig.~\ref{mocks_hrs} (lower middle panel, blue dots) candidates with $r_{SFR}\leq0.01$ show a better agreement between the input and output values of the mock with the exception of the very low values of $r_{SFR}$ that tends to be overestimated but still in the range of our criteria.
In addition to a selection from $r_{SFR}$, we impose that the galaxies must have more that one detection in IR to ensure a reliable L$_{IR}$ estimate from the IR SED.

With these two criteria, the final local sample of quenched galaxies is composed of 14 galaxies.
These galaxies are marked by the blue dots in Fig.~\ref{mocks_hrs}. 
For these galaxies, the $age_{\mathrm{trunc}}$ is very well constrained with all the points lying close to the one-to-one relationship, with the exception of the two galaxies with the shortest input quenching age, which tends to be overestimated.
This could be due to the known bias from the PDF analysis described earlier. 
In any case, for these two galaxies the error on the quenching age is large and will be taken into account in our analysis.
The SFR of the quenched candidates is well constrained except for the two sources mentioned above that have overestimated quenching age.
All the other physical properties of this quenched sample are well constrained according to the mock analysis.

These 14 galaxies are Virgo cluster members known to undergo ram pressure stripping.
This process removes the gas efficiently, especially in the outer parts of the disk, truncating the star formation activity outside-in \citep{Boselli06,Boselli16}.
The lack of both atomic and molecular gas \citep{Fumagalli09,Boselli14} reduces the star formation activity yielding to a migration of the galaxies from the blue cloud to the green valley and then the red sequence as predicted in models \citep{Boselli06,Boselli16}.
However, there have been cases of Jellyfish galaxies showing some enhancement of star formation despite undergoing ram pressure \citep{Durret21}. 
These galaxies have been selected in B or V optical bands from their Jellyfish morphology.
They thus show intermediate age stellar population in their morphological tails by selection.
Ram pressure stripping only affect gas, not stars, therefore it is probable that these galaxies underwent some gravitational interactions too, capable of removing stars as well as gas.
An increase of star formation due to ram pressure is only possible in the case of an almost edge-on interaction of a galaxy into the intra-cluster medium: the gas moves toward the disk before being remove out of the galaxy, this yields to a compression of the gas and thus a burst of star formation as observed in IC3476 \citep{Boselli21}.
This burst is relatively short as the stripping process is rapid ($<$500\,Myr), and is thus statistically difficult to observe.
Based on these arguments, we are confident that our selected HRS galaxies are indeed quenched due to ram pressure stripping, in agreement with observations and model predictions of this process.

\subsection{Comparison with the results from \cite{Boselli16}}

Using galaxies from the HRS sample, \cite{Boselli16} aimed at constraining the rapid decrease of the star formation activity of galaxies entering the dense environment of the Virgo cluster.
To do so, they combined UV to far-IR photometric data with age-sensitive Balmer absorption line indices extracted from medium-resolution (R$\sim$1000) integrated spectroscopy, as well as H$\alpha$ imaging data.
They use CIGALE combining all of their data and using a truncated SFH with a secular evolution parametrised using the chemo-spectrophotometric physically justified models of \cite{BoissierPrantzos00}.

Given the combination of spectroscopic (Balmer lines, including H$\alpha$) and photometric data (20 bands from UV to submm) and tailored SFHs, the results of \cite{Boselli16} are a reference and benchmark for ours.
From their SED modeling they obtained an estimate of the quenching age (QA) of their galaxies.
QA is the look-back time of the epoch of the quenching episode, which has the same definition than the $age_{trunc}$ parameter used in this work.
The two quantities can be thus directly compared.

\begin{figure}[!h] 
  	\includegraphics[width=\columnwidth]{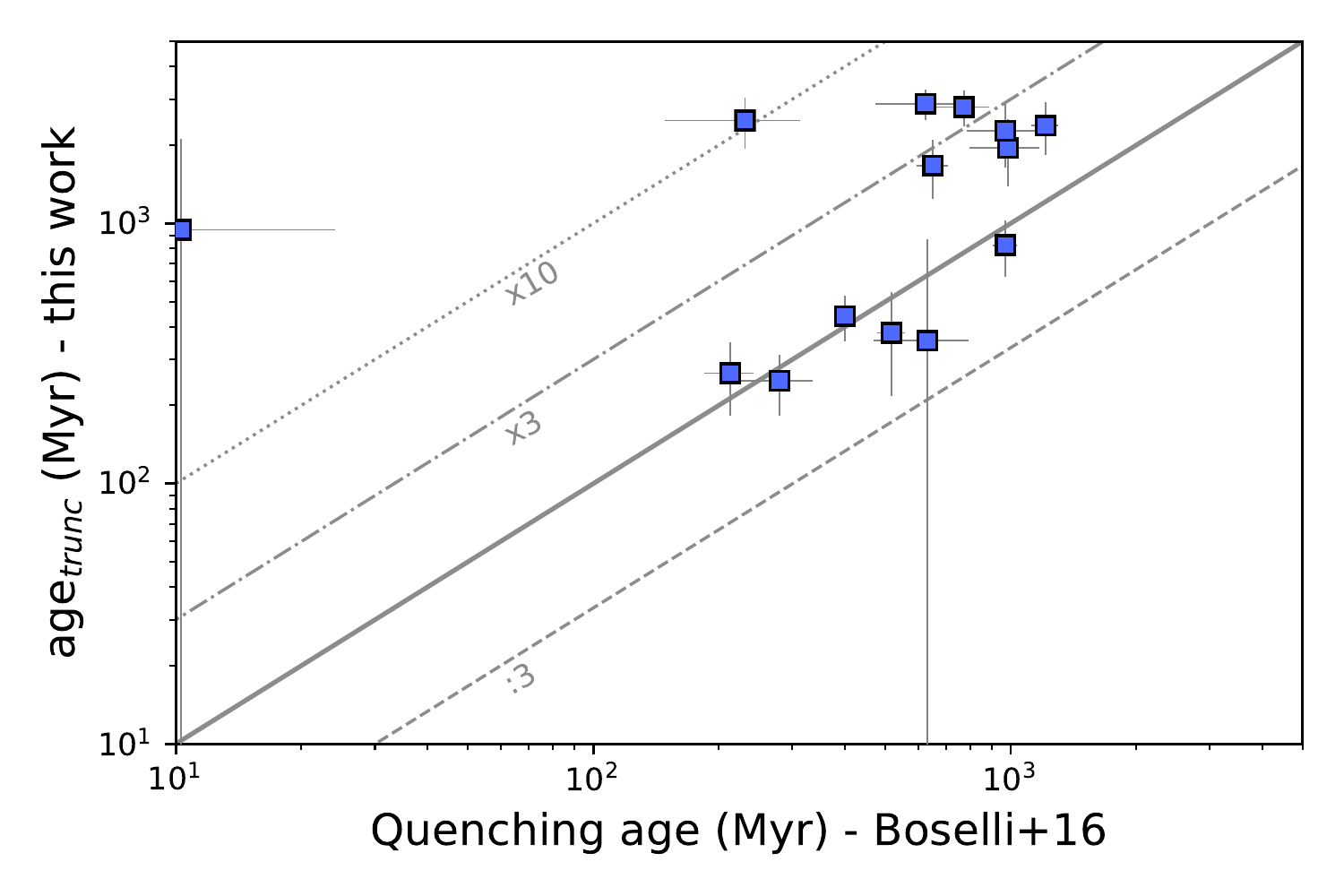}
  	\caption{\label{comp_al} Comparison between the age of quenching estimated in this work from broad band photometry and the same age estimated by \cite{Boselli16} from a combination of spectroscopic and photometric data. The solid line shows the one-to-one relationship.}
\end{figure}

To understand if our method, using only broad-band photometry allows us to recover sensible estimates of the age of quenching, we show in Fig.~\ref{comp_al} a comparison between our estimate of $age_{trunc}$ and the QA parameter of \cite{Boselli16}.
Even though the relationship is dispersed, same order of ages, that is within a factor of three, are found for 10 galaxies out of the 14.
Two galaxies are in strong disagreement with \cite{Boselli16} with our method estimating a quenching age between a factor 10 and higher than \cite{Boselli16}.
One of these two galaxies is found to give a very short quenching age by \cite{Boselli16} ($<20$\,Myr) whereas our method does not provide a good constraint on the value.
Considering the difficulty in estimating SFH parameters from broad-band SED fitting \citep[e.g.][]{Pforr12,Buat14,Ciesla15,Ciesla16}, the different attenuation law that we use, the good agreement within a factor 3 between the estimate of \cite{Boselli16} using spectroscopy information plus photometry and ours, we consider that our method does not introduce a strong bias in the quenching age determination.

\section{\label{cosmos}A complementary sample of 0.5$<z<$1 COSMOS quenched galaxies.}

The HRS sample is well-suited for our study as we have all the information needed regarding their UV to submm SED and the star formation quenching mechanism at play. 
However, it is a particular sample of galaxies at z$=$0 lying within the Virgo cluster.
Furthermore, the dynamical range probed by the HRS galaxies is quite limited in terms of luminosity.
We now want to understand if other galaxies follow the same relation or if the decrease of IR luminosity after quenching depends on other factors.
In the following, we make a first attempt to identify sources at higher redshift that could be used to complete our study, that is galaxies that just underwent a drastic and rapid decrease of their star formation activity, to be compared with the HRS selected galaxies.
A well-suited sample for this study is the COSMOS sample as it provides both a large wavelength coverage of the SED and the large statistics needed to pinpoint objects with very short variation of their SFH.

\subsection{COSMOS sub-sample, results from \cite{Aufort20}}

To select the high redshift galaxy sample, we rely on the results of \cite{Aufort20} who aimed at identifying galaxies having experienced a rapid and drastic variation of their star formation activity in the last 500\,Myr from a sample of COSMOS galaxies \citep[][]{Laigle16} with good quality data $S/N>$10.
They selected a sample of galaxies from the COSMOS sample with 0.5$<z<$1, stellar mass larger than 10$^{8.5}$\,M$_{\odot}$, and high S/N flux densities.
They developed a method based on Approximate Bayesian Computation \citep[ABC, see, e.g. ][]{marin2012approximate,sisson2018handbook} associated to a machine-learning algorithm \citep[XGBoost, see ][]{chen2016xgboost} to compute the probability that a galaxy experienced a recent, less than 500\,Myr, and drastic variation of its star formation activity that could be either an enhancement of the SFR or a quenching.
Based on the observed SED of a galaxy, they chose the most appropriate SFH between a finite set. 
The main idea behind ABC is to rely on many simulated SEDs generated from all the SFHs in competition using parameters drawn from the prior. 
For each galaxies of their sample, the posterior probability $p$ that a galaxy has experienced a recent and rapid variation of star formation activity is computed.
We rely on this probability to select our galaxies and conservatively select those with $p>0.91$ which corresponds to galaxies where there is a \textit{very strong} to \textit{decisive} evidence for a recent strong variation of the SFH according to the Jeffreys scale \citep[see, e.g., ][]{Robert07}.
Out of the 12,380 galaxies of their sample, 376 galaxies have a posterior probability higher than 0.91.

Although the results of \cite{Aufort20} allows us to select galaxies with a recent variation of SFH, their method does not provide information on the nature of this variation, that is if a galaxy underwent a starburst phase or a quenching of SF.
For the purpose of this study, we need to select galaxies that underwent a strong decrease of SFR.
Therefore we combine UV to IR data of our COSMOS sub-sample of 376 galaxies from the catalogues of \cite{Laigle16} and \cite{Jin18} to determine the nature of the variation from SED modelling.
We include the intermediate bands as well as all the \textit{Spitzer}/IRAC ones.
We use the \cite{Jin18} IR COSMOS catalogue from 24\,$\mu$m to 350\,$\mu$m (no detection is found at longer wavelengths for the 376 galaxies).
In IR, we only consider detection with a S/N larger than 3.
The list of bands used in this study is provided in Table~\ref{bands} as well as the number of detection in each of them.

\begin{table}
	\centering
	\caption{COSMOS broad and intermediate bands used in this work. The data up to \textit{Spitzer}/IRAC4 are from \cite{Laigle16} while the IR data are from \cite{Jin18}. The number of sources with a detection in each bands is provided in the last two columns for the whole sample and the selected quenched sources.}
	\begin{tabular}{l c c c c}
	 \hline\hline
	Instrument & Band & $\lambda$ ($\mu$m) & \# of sources & \# final \\ 
	\hline
	GALEX            & FUV & 0.153 & 331  & 5\\
	GALEX            & NUV & 0.229 &  376  & 7\\
	CFHT        & $u'$ & 0.355 & 376   & 7\\
	SUBARU          & B & 0.443&  376  & 7\\
	SUBARU         & V & 0.544& 376 & 7\\	   
	SUBARU         & r & 0.622&376  & 7\\	 
	Suprime Cam    & $i'$ & 0.767& 376 & 7\\	
	Suprime Cam    & $z'$ & 0.902& 376 & 7\\	
	Intermediate bands &  & 0.427  & 376 & 7\\
	      "              &  & 0.464  & 376 & 7\\
	      "              &  & 0.505 & 376 & 7\\
	   "             &  & 0.527 & 376 & 7\\
	       "             &  & 0.574 & 376 & 7\\
	       "             &  & 0.624 & 376 & 7\\
	       "             &  & 0.709 & 376 & 7\\
	       "             &  & 0.738 & 376 & 7\\
	       "             &  & 0.767& 376 & 7\\
	       "             &  & 0.827& 376 & 7\\
	HSC            & Y &  & 376 & 7\\	  
	VISTA            & Y & 1.019& 376 & 7\\	
	VISTA            & J & 1.250& 376 & 7\\	
	VISTA            & H & 1.639& 376 & 7\\	
	VISTA            & Ks & 2.142& 376 & 7\\	
	\textit{Spitzer} & IRAC1 & 3.6& 376 & 7\\	
	\textit{Spitzer} & IRAC2 & 4.5& 376 & 7\\	
	\textit{Spitzer} & IRAC3 & 5.8& 300 & 5\\	
	\textit{Spitzer} & IRAC4 & 8.0& 179 & 1\\
	\textit{Spitzer} & MIPS & 24& 194 & 4\\	
	\textit{Herschel} & PACS & 100& 53 &  0\\
	\textit{Herschel} & PACS & 160& 40 & 0\\
	\textit{Herschel} & SPIRE & 250& 42 & 1\\
	\textit{Herschel} & SPIRE & 350& 18&  1\\
	\hline
	\label{bands}
	\end{tabular}
\end{table}

\subsection{Constraints on the parameters for the COSMOS selected galaxies}
\begin{figure*}[!h] 
  	\includegraphics[width=\textwidth]{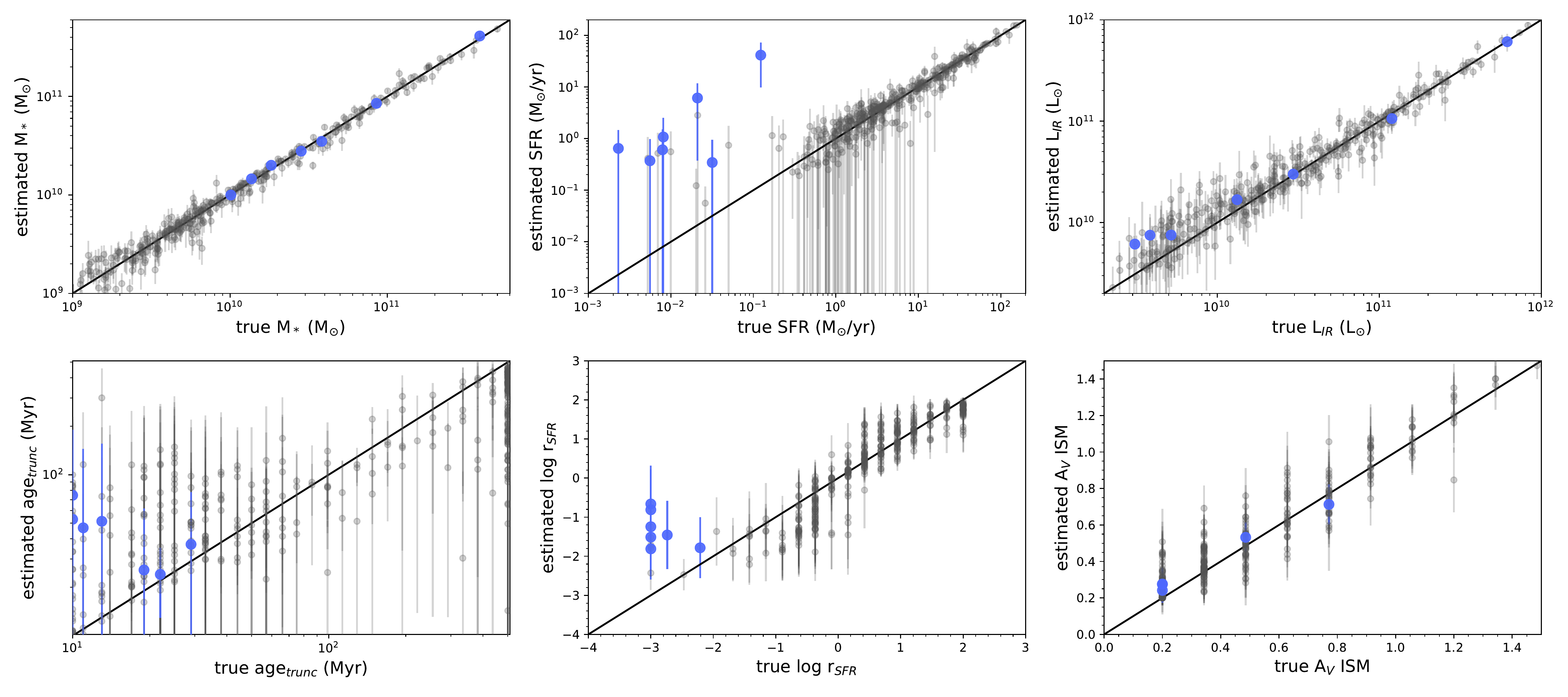}
  	\caption{\label{mocks} Results of the mock analysis. The input parameters used to build the mock catalogue are shown on the x-axis while the results of the fitting of the mock catalogues are shown on the y-axis. A good constraint on a given parameter is obtained when there is a one-to-one relationship, which is indicated by the black solid lines. Grey dots are all sources selected from the \cite{Aufort20} sample while blue circles are the galaxies composing the final sample.}
\end{figure*}

As for the HRS galaxies, we proceed to a mock analysis to understand how well the output parameters from the fit are constrained.
Indeed, given the redshift of the SED as well as the different photometric bands used compared to HRS, the results obtained for the local galaxies are not necessarily applicable in the case of the COSMOS galaxies.
The results of the mock analysis performed on our sample of 376 galaxies is shown in Fig.~\ref{mocks}.
Similarly to the HRS galaxies, the stellar mass, SFR, L$_{IR}$, and A$_V$ ISM attenuation are well constrained. 

Regarding the parameters linked to the SFH, the relation between the input value and the one recovered by CIGALE of the $r_{\mathrm{SFR}}$ parameter does not follow exactly the one-to-one relationship for the same reason than the one explained above for the HRS galaxies (bias due to the PDF analysis).
Nevertheless, this known bias is not a strong issue for our analysis.
Indeed, as seen from Fig.~\ref{mocks}, the recovered $r_{\mathrm{SFR}}$ values from input $\log r_{\mathrm{SFR}} < 0$ remains below 0.
This means that selecting galaxies with a negative value of $\log r_{\mathrm{SFR}}$ is a conservative approach as this value can be slightly overestimated due to the PDF analysis performed by CIGALE.

The $age_{\mathrm{trunc}}$ parameter shows a more dispersed relation between the input and output value with large uncertainties.
However, as we will discuss in the following section, the selection criteria that we will apply allows us to be relatively confident on the estimate of $age_{\mathrm{trunc}}$ of our quenched candidates. 

\subsection{Selection of recently quenched galaxies}
To separate galaxies having experienced a recent starburst from those that have been quenched, we run CIGALE on the 376 galaxies that have a probability higher than 91$\%$ according to \cite{Aufort20}.
The input parameters used in CIGALE are the same than for the mock analysis and are provided in Table~\ref{inputparam}.
However, to limit degeneracies, we use an option of CIGALE which is the possibility to provide parameters to be fitted in the same way than any other photometric flux density.
We thus assume an age of the COSMOS galaxies, based on their redshift, and put it as an input to be fitted by the code.
Therefore it is not a fixed parameter but a strong constraint for the SED fit.
This reasonable assumption allows to reduce the free parameters of the SFH modelling from four to three.

To be conservative, we apply to the COSMOS galaxies the same selection criteria than for the HRS local sample, that is $r_{\mathrm{SFR}}\leq0.01$.
Out of the 376 galaxies of our sample, 7 satisfy this criteria.

The constraints on the parameters obtained from SED fitting for these sources are shown in Fig.~\ref{mocks} in blue.
The SFR of these galaxies is more uncertain and slightly overestimated compared to the other galaxies of the full sample of 376 sources which is not surprising given the low star formation activity of these sources.
The bias in the extreme values of the mock analysis may play a role into this too.
However, as explained in Sect.~\ref{cons}, our selection is conservative as the actual $r_{\mathrm{SFR}}$ and thus SFR values may be lower than what is estimated by CIGALE.
Regarding the age of quenching, $age_{\mathrm{trunc}}$, the PDF analysis of the parameter can yield to an overestimation of the true value by a factor of 4, at most.
The $age_{\mathrm{trunc}}$ of quenched candidates with an estimated $age_{\mathrm{trunc}}$ lower than 50\,Myr are constrained since the output age resulting from the mock analysis is below 50\,Myr too.
Although we are using broad-band SED fitting to estimate variations of SFH on very short time scales ($<100$\,Myr), the UV rest-frame data are sensitive enough to be able to probe these scales with SED fitting as demonstrated in \cite{Boquien14} using hydro-dynamical simulations of main sequence galaxies.
Despite the large errors, we can still have an information on $age_{\mathrm{trunc}}$ which remains short ($<100$\,Myr). 

As a sanity check, we run CIGALE on these 7 galaxies using only a normal delayed-$\tau$ SFH and compare the quality of the fits provided by the two models (normal delayed-$\tau$ and flexible SFH).
As their SFR is very low, we want to be sure that their SED could not be fitted by a normal delayed-$\tau$ SFH with low values of $\tau_{main}$ ($<3\,Gyr$) which is usually assumed to model passive and quiescent galaxies.
To do so, we compute the Bayesian Information Criterion (BIC) for each SFH assumption and calculate the difference between them, that is $\Delta$BIC \citep[see][for more details]{Ciesla18,Buat19,Aufort20}.
For the 7 galaxies of our final sample, $\Delta$BIC is larger than 10 which is the threshold to claim that the evidence against a normal delayed-$\tau$ SFH is decisive \citep[see, e.g., ][]{Robert07}.
This strengthens the results of the \cite{Aufort20} method in selecting galaxies with a recent and strong variation of SFH and ensure that our sample is not contaminated by passive and smoothly quenched galaxies.

To confirm if our selection yields a sample of quenched galaxies, we search for optical spectra for the 7 galaxies.
We find three galaxies with a zCOSMOS \citep{Lilly09} optical spectra, that we retrieved from the ASPIC\footnote{\url{http://cesam.lam.fr/aspic/}} database, shown in Appendix~\ref{optspec}.
The optical spectra of the three galaxies show no strong emission lines that could be hints of a star formation activity.
It indicates that $age_{trunc}$ is larger than 10\,Myr, the typical age of HII regions.
These three sources are thus confirmed to be quenched and serve as fiducial indicators in the following.

Out of the 12,380 galaxies from the sample of \cite{Aufort20}, we select 7 galaxies. 
However, we can not deduce any statistical information out of this number on the population of galaxies undergoing rapid quenching.
Indeed, first of all, in the selection of their sample  \cite{Aufort20} adopted some criteria to keep the statistical problem simple, such as a S/N cut and a detection of the galaxies in all the main photometric bands they used.
Therefore their sample is not complete.
Furthermore, the present study being a first attempt in recovering the past recent properties of galaxies undergoing a rapid quenching of their SF, we are very conservative in our criteria in order to have a clean sample. 
Here, again, our selection do not provide a complete sub-sample of recently quenched galaxies.

\subsection{Physical properties of the selected quenched galaxies}
\begin{figure*}[!h] 
  	\includegraphics[width=6.3cm]{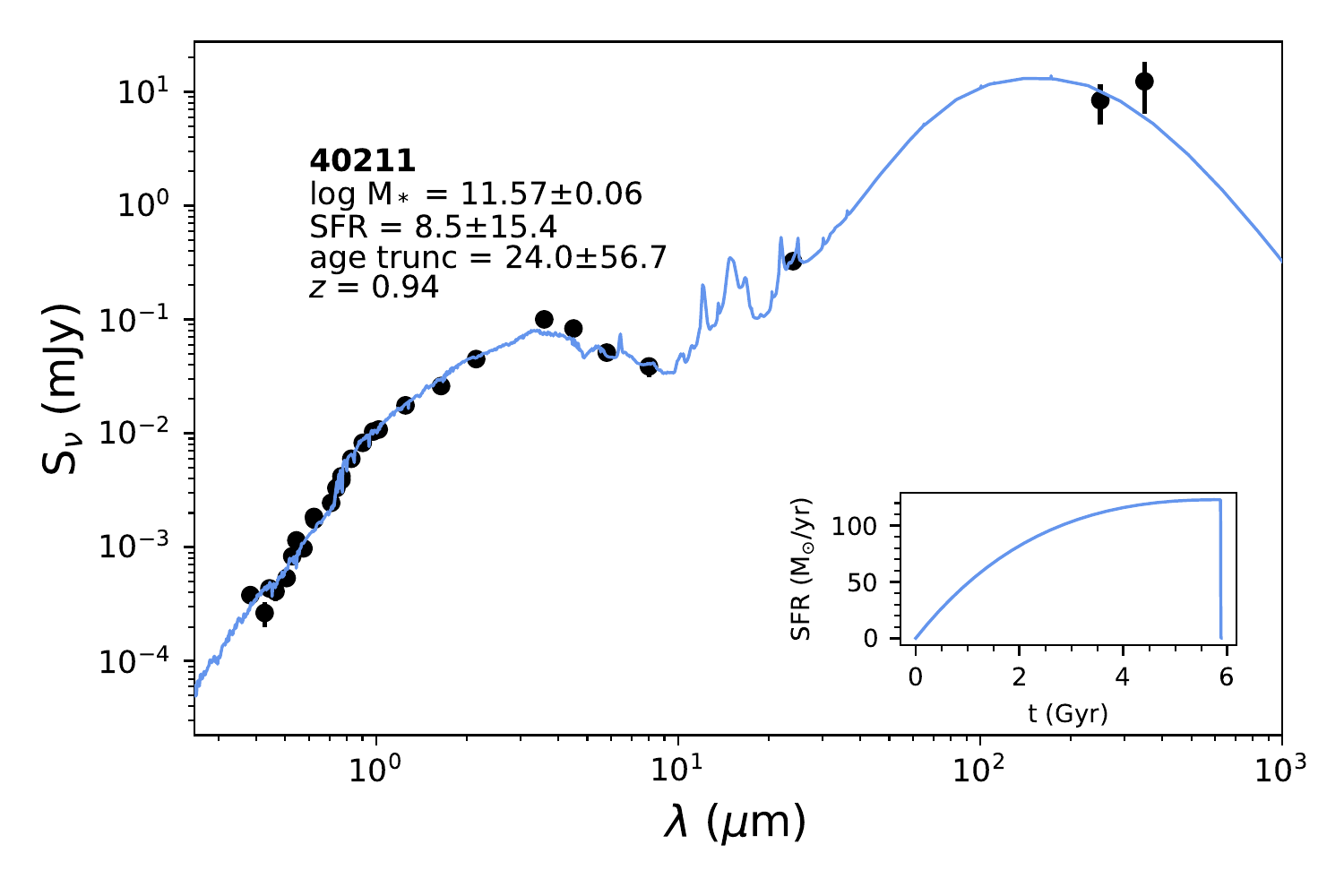}
  	\includegraphics[width=6.3cm]{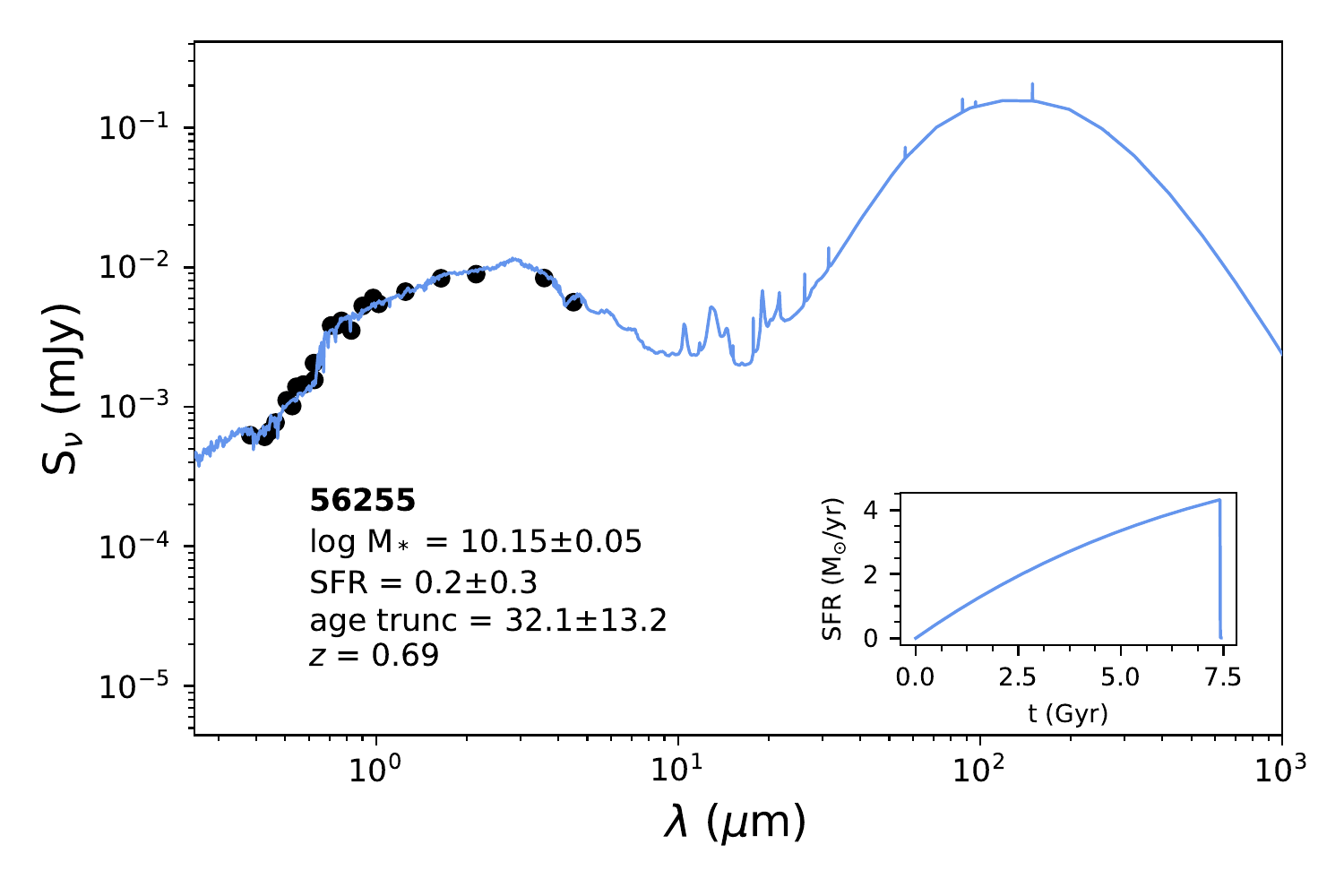}
  	\includegraphics[width=6.3cm]{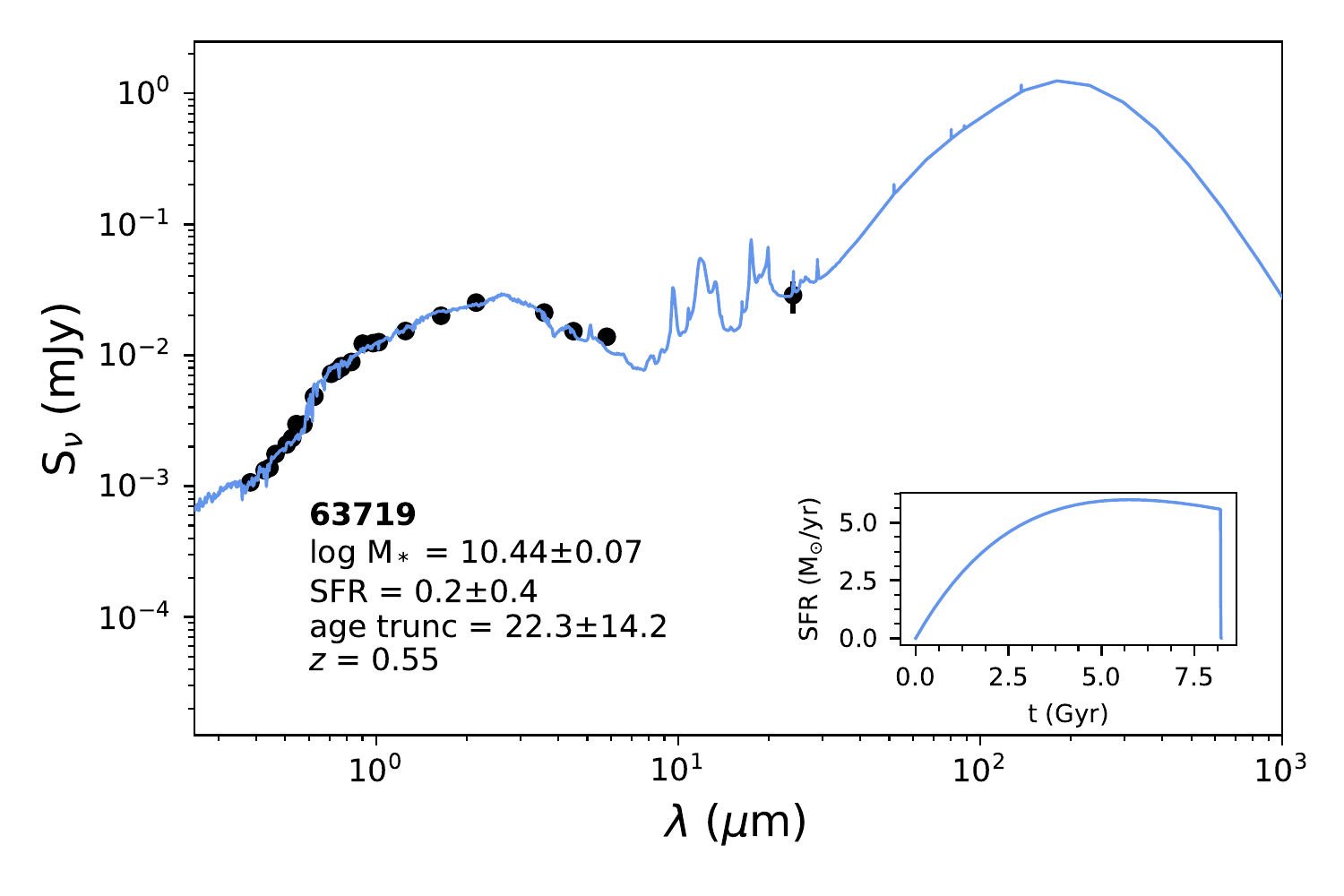}
  	\caption{\label{SEDs} Examples of SEDs of galaxies from the final COSMOS quenched sample fitted with CIGALE. The black dots are the data points while the blue solid lines indicate the best fit model. The inset panels show the SFH obtained from the best fit. Key physical parameters are provided for each galaxy. These three galaxies are those confirmed as quenched from their optical spectra.}
\end{figure*}

Examples of the fits obtained by CIGALE for the three spectroscopically confirmed quenched galaxies are shown in Fig.~\ref{SEDs}.
The stellar masses, IR luminosities, attenuation (A$_V$), and quenching ages of each of the 7 candidates are shown in Table~\ref{properties}.

\begin{table*}
	\centering
	\caption{Physical properties obtained with CIGALE for the 7 quenched candidates of the final high-$z$ sample.}
	\begin{tabular}{l c c c c c c c}
	 \hline\hline
	id & redshift & 10$^{10}M_*$ (M$_{\odot}$) &   10$^{9}L_{IR}$ (L$_{\odot}$) & A$_V$ & $age_{\mathrm{trunc}}$ & Optical spectra & MIPS detected\\ 
	\hline
	5043   & 0.63  &1.01$\pm$0.08   & 7.5$\pm$5.6   &  0.34$\pm$0.13 & 42$\pm$89 & NO & NO\\
	40211  & 0.94  &37.34$\pm$2.41  &  603.2$\pm$77.9  &  2.15$\pm$0.14 & 24$\pm$57 & YES& YES\\
	43190  & 0.59  &3.76$\pm$0.38   &  30.3$\pm$4.7  &  0.75$\pm$0.11 & 49$\pm$99 & NO & YES\\
	51525  & 0.74  &8.37$\pm$0.81   &  122.7$\pm$15.1  &  2.11$\pm$0.14 & 54$\pm$94 & NO & NO\\
	56255  & 0.69  &1.40$\pm$0.07   &  4.5$\pm$1.2  &  0.21$\pm$0.03 & 32$\pm$13 & YES& NO\\
	63719  & 0.55  &2.74$\pm$0.18   &  15.2$\pm$3.0  &  0.53$\pm$0.08 & 22$\pm$14 & YES& YES\\
	74332  & 0.53  &1.87$\pm$0.09   &  6.0$\pm$1.9  &  0.21$\pm$0.04 & 25$\pm$8  & NO & NO\\	  
	\hline
	\label{properties}
	\end{tabular}
\end{table*}

\begin{figure}[!h] 
  	\includegraphics[width=\columnwidth]{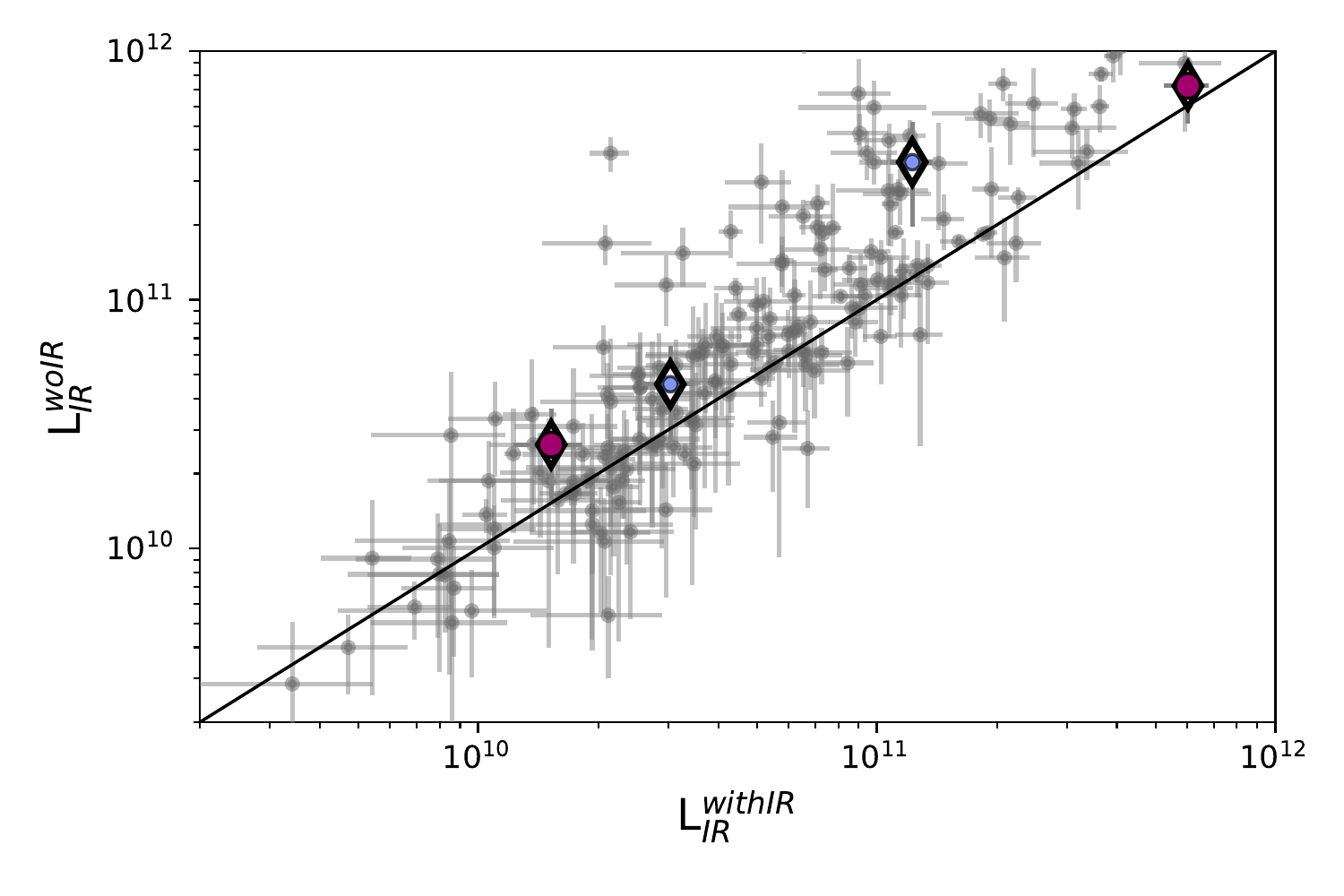}
  	\caption{\label{comp_lir} Comparison between the L$_{IR}$ obtained by CIGALE from UV-IR SED modelling with and without using the MIPS 24\,$\mu$m flux density for galaxies which are MIPS detected. Candidates of the final sample with a MIPS 24\,$\mu$m detection are indicated by the diamonds. Red circles are the candidates of the final sample for which the quenching is confirmed by spectroscopy.}
\end{figure}

In our final sample, only 4 of the galaxies have an IR detection.
To check the validity of the L$_{IR}$ estimate by CIGALE in absence of such measurement for the three remaining, we compare in Fig.~\ref{comp_lir} the L$_{IR}$ obtained by CIGALE with and without using the available MIPS 24\,$\mu$m flux density.
For this test we use galaxies of the initial sample of 376 sources that are detected in MIPS 24\,$\mu$m.
There is a relatively good one-to-one relationship between the two measurement, especially since we consider here galaxies that have either undergone a star-bursting event or a quenching of their star formation activity, in other words galaxies outside the galaxy star-forming main sequence.
This is consistent with the results of \cite{Malek18} who performed the same test on a large sample of IR galaxies from HELP\footnote{\url{https://herschel.sussex.ac.uk/}} \citep[\textit{Herschel} Extragalactic Legacy Project][]{Vaccari16} and found a good relation between the IR luminosity estimates with and without using the IR data. 
This relation allows us to consider that the L$_{IR}$ estimated by CIGALE in the absence of IR data is a fair approximation for the true L$_{IR}$ of the galaxy, that we will use in the rest of the paper.

\section{\label{evol}The evolution of IR luminosity after quenching}

We now combine the local and high redshift samples.
As discussed above, directly from the modelling of the UV-submm SED, we have an estimate of the time at which the quenching of star formation happened, $age_{\mathrm{trunc}}$.
From the fit of the IR SED, we have measured the L$_{IR}$ at the time the galaxy is observed.
To investigate the evolution of the IR luminosity after the shutdown of the star formation activity, we need to estimate the IR luminosity just before the quenching as a reference to quantify the decrease since the quenching.
We thus need to recover the past star formation activity, traced by the L$_{IR}$, of our quenched candidates.
To have an estimate of this, for each galaxy of the joined sample (HRS+COSMOS), we use the SFH best fit parameters of the observed UV-submm SED and build the SED just before quenching.
In details, we use the long-term SFH parameters ($\tau_{main}$ and $age$). 
By building the past SED, we determine the L$_{IR}$ at the time just before quenching, that we call L$_{IR}^{bq}$ for ``before quenching''.

According to the hypothesis driven by our SFH model, our quenched candidates are supposed to have been normally forming stars and then abruptly quenched their star formation activity.
If this is indeed the case, then the recovered past IR luminosity, L$_{IR}^{bq}$, should be consistent on average with the L$_{IR}$ of a reference sample of normal star-forming galaxies with the same stellar mass\footnote{We checked that the quenching does not affect significantly the stellar mass of the galaxies.}.
For each local and high-redshift candidate, we build a reference sample of galaxies with stellar masses between 0.8 and 1.2 the stellar mass of the candidate and compute the median L$_{IR}$ of this reference sample, L$_{IR}^{ref}$.
For the HRS quenched galaxies, the bins are drawn from from the whole sample of 260 late-type HRS galaxies, and using only the IR SED fit.
For the COSMOS quenched galaxies, the bins are drawn from the whole initial sample of \cite{Aufort20} of 12,380 galaxies, and using only the L$_{IR}$ obtained from the UV-submm SED fit as we showed that it provides a good estimate of the true L$_{IR}$ even with sparse IR sampling of the SED.

\begin{figure}[!h] 
  	\includegraphics[width=\columnwidth]{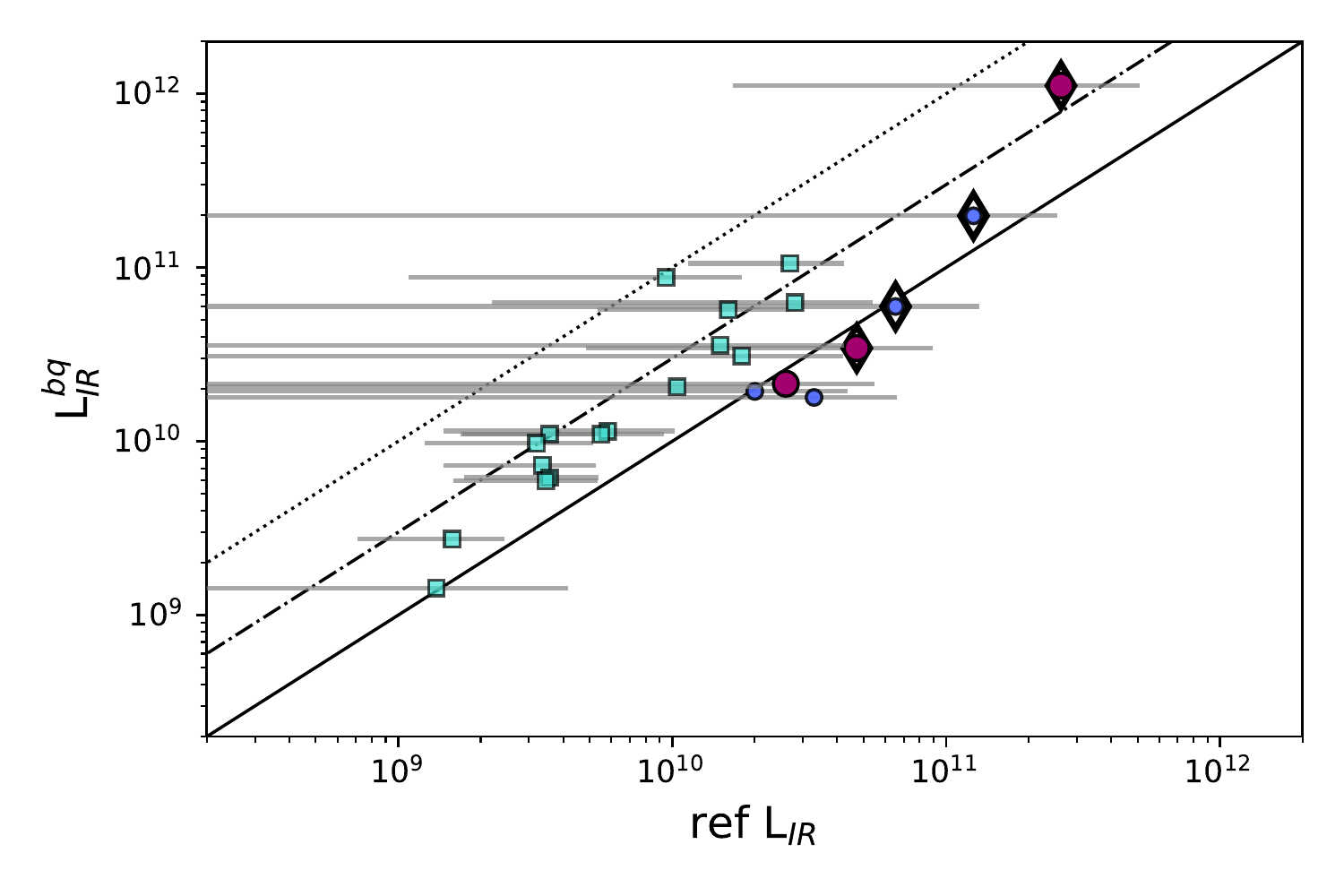}
  	\caption{\label{lbq_lref} Estimated L$_{IR}$ of the sources before quenching as a function of the median L$_{IR}$ of a reference sample with similar stellar mass. The cyan squares are the HRS quenched galaxies while the dots are the COSMOS ones. COSMOS galaxies with a diamond are sources having a 24$\mu$m detection. Red dots are COSMOS quenched galaxies that are spectroscopically confirmed. The error bars show the dispersion of the L$_{IR}$ distribution of each reference sample. The solid line shows the one-to-one relationship while the dashed-dotted line a factor three above it, and the dotted line a factor 10 above it.}
\end{figure}

\begin{figure}[!h] 
  	\includegraphics[width=\columnwidth]{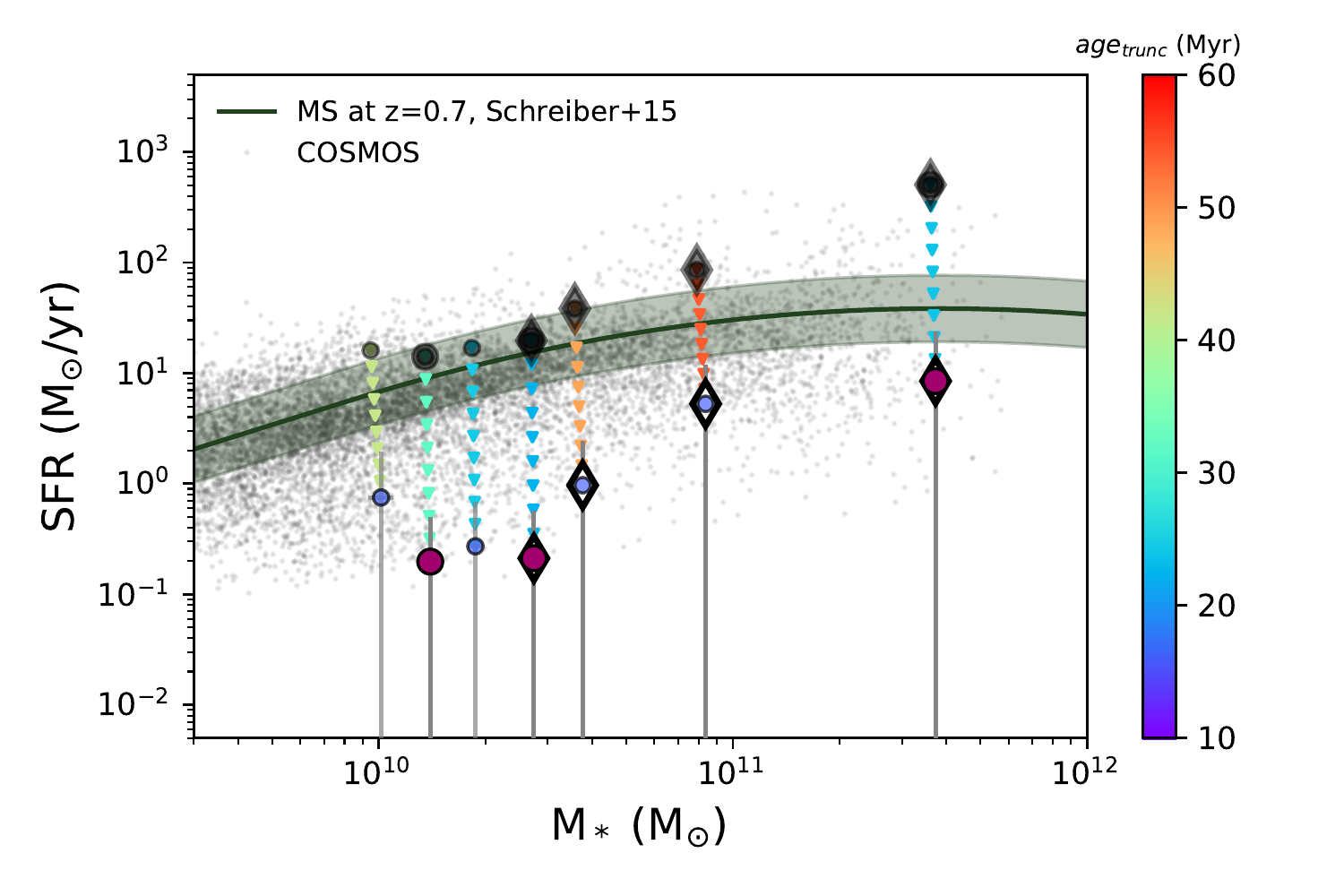}
  	\caption{\label{ms} Star formation rate as a function of stellar mass for the COSMOS galaxies. Grey dots are the whole COSMOS sub-sample of \cite{Aufort20}. Blue dots, red dots, diamonds, are the 7 COSMOS quenched sources studies in this work. Their position relative to the star-forming main sequence of galaxies before their quenching is marked by grey symbols. Their evolution since quenching is indicated by the triangles coloured according to their $age_{\mathrm{trunc}}$. As an indication the main sequence of \cite{Schreiber15} at $z\sim0.7$ is shown in solid black line along with its dispersion (shaded grey region). }
\end{figure}

In Fig.~\ref{lbq_lref}, we show the L$_{IR}^{bq}$ of each HRS and COSMOS quenched galaxies as a function of their corresponding L$_{IR}^{ref}$.
Six out of the seven COSMOS galaxies are lying very close to the one-to-one relationship.
We have checked that the large error on $age_{\mathrm{trunc}}$ for four out of 7 galaxies of COSMOS does not impact the estimate of L$_{IR}^{bq}$ by varying $age_{\mathrm{trunc}}$ within the error and found that our measurement is stable.
For the COSMOS sources, we are able with broad-band SED fitting to recover the L$_{IR}$ before the quenching of star formation.
This may be due to the fact that these COSMOS galaxies quenched recently, that is less than 100\,Myr ago according to our estimate of $age_{\mathrm{trunc}}$.
However, for one galaxy (\#40211) the L$_{IR}^{bq}$ seems to be larger than the corresponding L$_{IR}^{ref}$ by a factor of three approximately.
This galaxy benefits from good data as it has been detected in 24$\mu$m and its quenching is confirmed from optical spectroscopy.
Given the fact that our method provides good estimates of the L$_{IR}$ before quenching of the six other COSMOS sources, one could interpret the discrepancy between the L$_{IR}^{bq}$ of \#40211 and its corresponding L$_{IR}^{ref}$ as possible indication that this galaxy was experiencing a star-bursting phase just before quenching, hence the high L$_{IR}^{bq}$.
Following a similar approach than in \cite{Ciesla18}, we show in Fig.~\ref{ms} the present and past positions of the 7 COSMOS quenched sources on the MS diagram.
Six out of the 7 are compatible with lying on or close to the MS before their quenching.
The seventh source, \#40211, seems to have been in a star-bursting phase before undergoing a star formation activity decrease as we just discussed.
This diagram indicates the relative short timescale with which the COSMOS galaxies have taken off from the MS providing tentative indications on galaxies movements within the MS.
This is consistent with the tight scatter observed in the MS implying that variations of the star formation activity should happen on relatively short time scales \cite[e.g.][and references therein]{ForsterSchreiber20}.

\begin{figure}[!h] 
  	\includegraphics[width=\columnwidth]{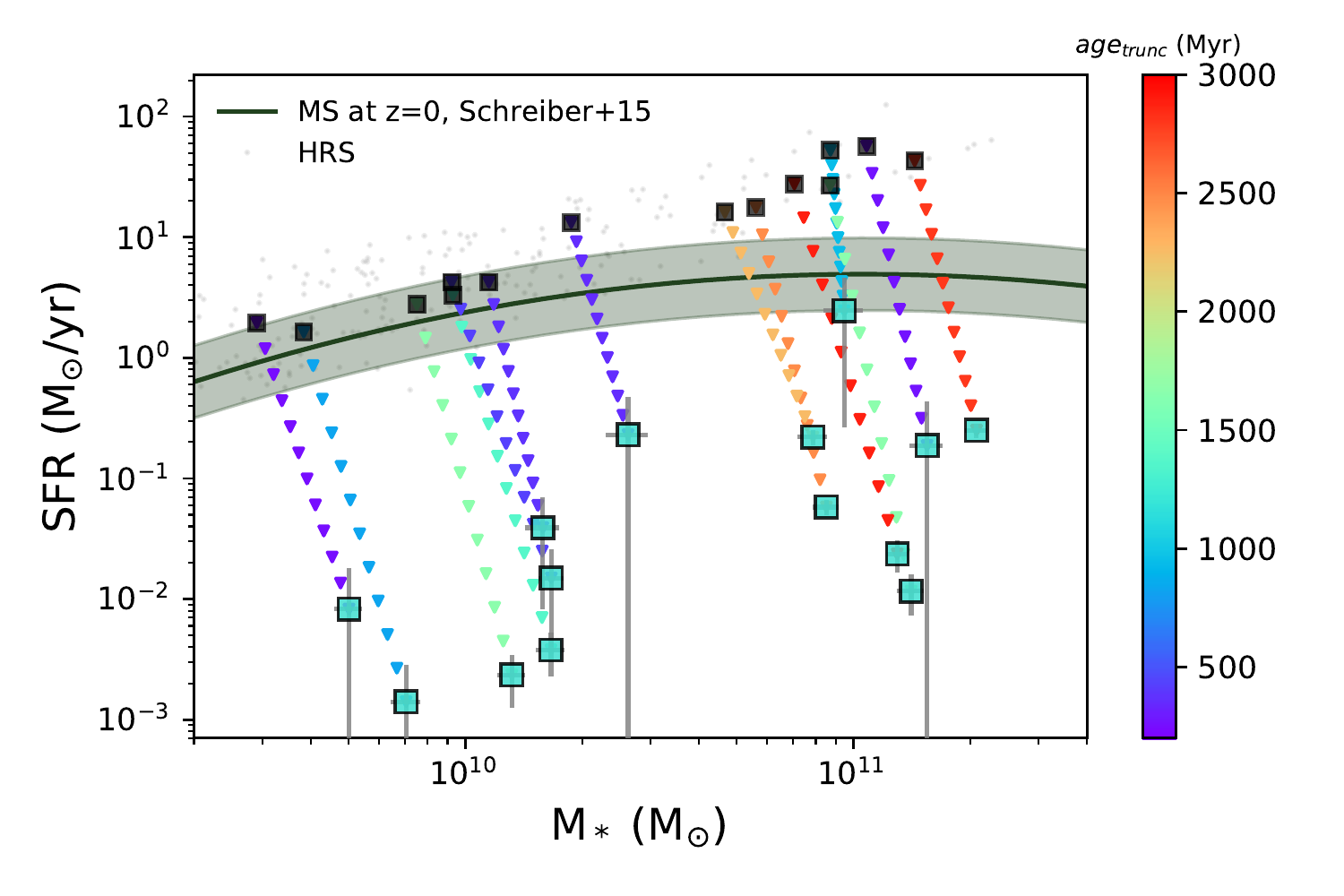}
  	\caption{\label{ms_HRS} Star formation rate as a function of stellar mass for the HRS galaxies. Grey dots are late-type HRS galaxies. The squares are the HRS quenched sources studies in this work. Their position relative to the star-forming main sequence of galaxies before their quenching is marked by grey symbols. Their evolution since quenching is indicated by the triangles coloured according to their $age_{\mathrm{trunc}}$. As an indication the main sequence of \cite{Schreiber15} at $z=0$ is shown in solid black line along with its dispersion (shaded grey region). }
\end{figure}

Regarding the HRS quenched galaxies, 11 out of the 14 candidates have L$_{IR}^{bq}$ and L$_{IR}^{ref}$ consistent within a factor between 1 and 3.
For these galaxies, the probed $age_{\mathrm{trunc}}$ are longer with values between 300\,Myr to 3\,Gyr.
These longer timescales make the recovering of the past SED more challenging.
The HRS is well-known galaxy sample for which a wealth of ancillary data and studies is available and we know that in this case the slight overestimation of the L$_{IR}^{bq}$ is probably not due to a past star-bursting phase.
However, although they do not lie right on the one-to-one relationship, our method is able to recover the IR luminosity, and thus star formation property, of the galaxies within a factor of 3 on timescales of a few hundreds of Myr to a couple of Gyr.
As for the COSMOS quenched sources, we place the HRS quenched sources on the MS diagram (Fig.~\ref{ms_HRS}) to recover their position before quenching.
They were compatible with the $z=0$ MS of \cite{Schreiber15}.
The galaxies with the larger values of $age_{trunc}$ (larger than a couple of Gyr) are the most massive sources. 

\begin{figure}[!h] 
  	\includegraphics[width=\columnwidth]{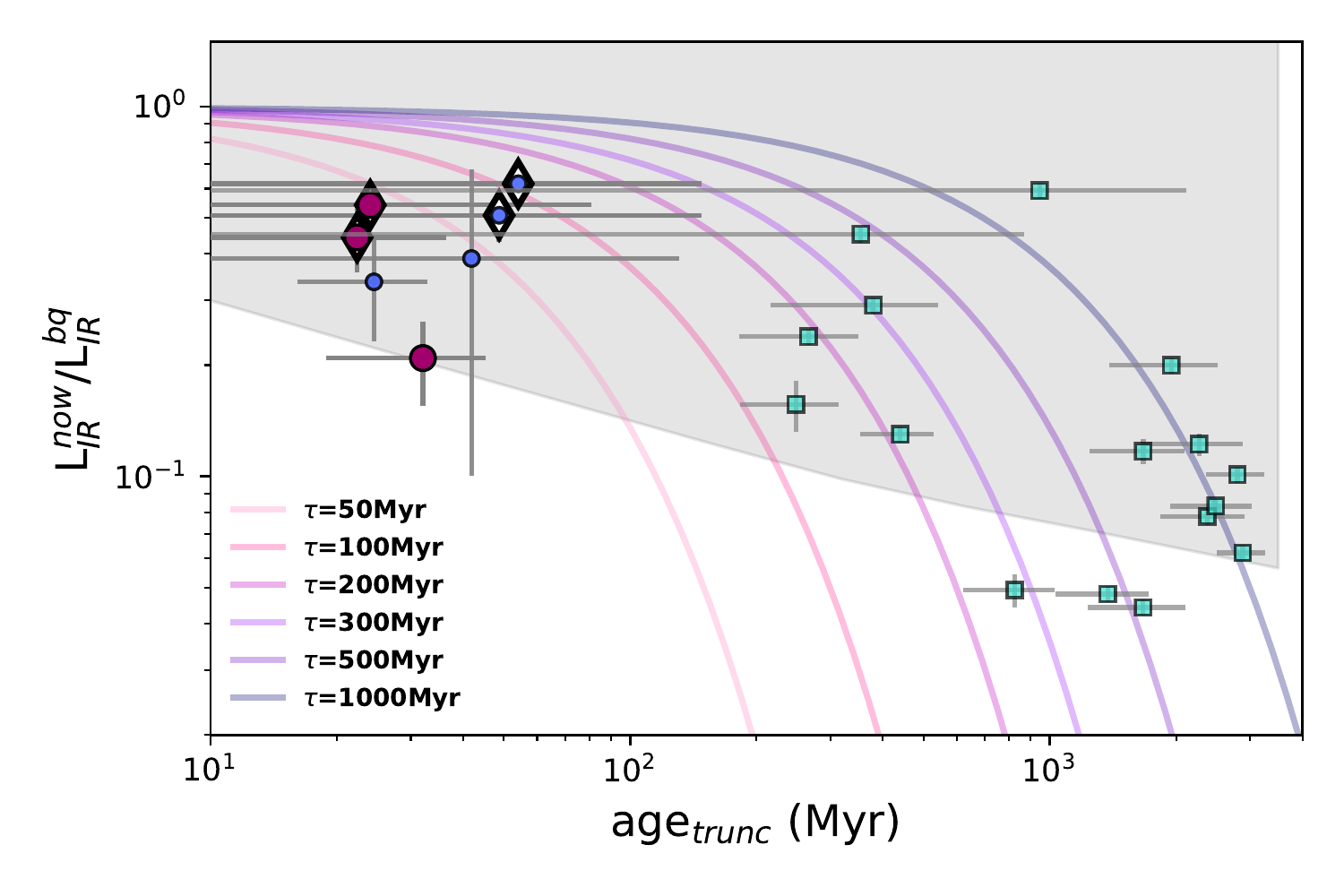}
  	\caption{\label{lir_vs_agetrunc_all} Ratio between the observed L$_{IR}$ and the L$_{IR}$ before the truncation of the SFH as a function of the age of truncation of the SFH ($age_{\mathrm{trunc}}$). Circles are galaxies from the COSMOS quenched sample, the red ones are galaxies confirmed from their spectra. Diamonds indicate galaxies for which IR data is available. Cyan squares are galaxies from the local HRS quenched galaxies. The light grey shaded region indicates the region covered by the UV-NIR SED models used to determine the physical properties of the galaxies. The coloured tracks indicate an exponential decrease of L$_{IR}^{now}$/L$_{IR}^{bq}$ as a function of age$_{\mathrm{trunc}}$ assuming different e-folding times.}
\end{figure}

We show the L$_{IR}^{now}$/L$_{IR}^{bq}$ ratio, that are the current L$_{IR}$ and the one before quenching, respectively, as a function of $age_{\mathrm{trunc}}$ in Fig.~\ref{lir_vs_agetrunc_all} for all the HRS+COSMOS quenched galaxies.
The two samples are complementary as the COSMOS galaxies probe shorter $age_{\mathrm{trunc}}$ compared to the local galaxies complementing the dynamical range in $age_{\mathrm{trunc}}$.
To interpret the position of the sources on this diagram, we add tracks assuming an exponential decrease of the L$_{IR}^{now}$/L$_{IR}^{bq}$ ratio as a function of time after quenching assuming different e-folding times $\tau$ (from 50 to 1,000\,Myr).
The position of the COSMOS galaxies, despite their large errors on the $age_{\mathrm{trunc}}$ is compatible with a decrease with a short e-folding time that is less than 300\,Myr.
Out of the 14 HRS quenched galaxies, five are also compatible with these short $\tau$ values of a couple of hundreds of Myr.
Two HRS galaxies have a very uncertain estimates of $age_{\mathrm{trunc}}$ that do not allow us to discuss their position of the diagram.
However, the position of the other HRS quenched galaxies are compatible with a decrease with a longer timescale, seven of them are lying on the track of a decrease with an e-folding time of 1\,Gyr. 
This is consistent with the physical process that caused their star formation quenching which ram pressure stripping due to the environment of the Virgo cluster.
The positions of the COSMOS quenched sources on Fig.~\ref{lir_vs_agetrunc_all} imply a rapid and drastic physical process.
Six out of the seven candidates seem to be compatible with a process linked to short timescales lower than $\sim$100\,Myr.
Star formation fluctuations on this timescale can be due to formation and destruction of individual giant molecular clouds where feedback is locally too weak to prevent gravitational collapse \citep[e.g.][, and references therein]{ScaloStruckMarcell84,FaucherGiguere18,Orr19,Tacchella20}.
However, the errors on age$_{\mathrm{trunc}}$ makes the processes compatible with longer timescales of the order of a few hundreds of Myr.
In this time ranges, star formation can be affected by galaxy mergers, bar-induced inflows, disk instabilities, galactic winds or environmental effects \citep{GunnGott72,Hernquist89,MihosHernquist96,Robertson06,OppenheimerDave08,McQuinn10,DekelBurkert14,Zolotov15,Tacchella16,Sparre17,Torrey18,WangLilly19,Tacchella20}.
Further investigation on each of the COSMOS candidate is needed to identify the process at play.

\section{\label{discussion}Discussion}

In the previous section, we put in evidence a decrease of the IR luminosity with different timescales but we did not put constraints on the origin of the decrease.
Is it due to the absence of young stars heating the dust? does a lack of dust content contributes as well?

\begin{figure}[!h] 
  	\includegraphics[width=\columnwidth]{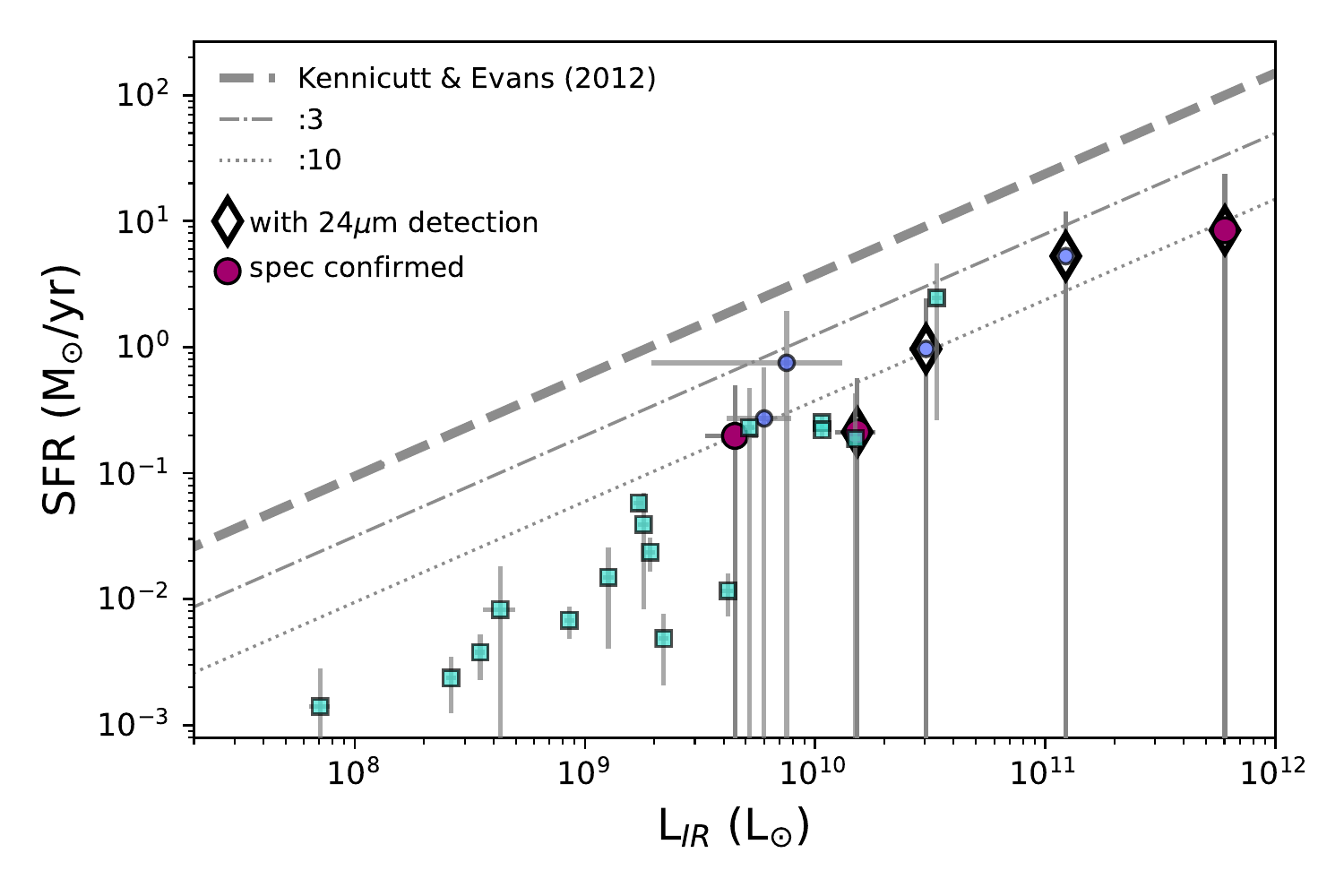}
  	\caption{\label{lirsfr} SFR as a function of IR luminosity for the galaxies of the final sample obtained from the UV-submm SED fitting. Circles are galaxies from the final COSMOS sample, the red ones are galaxies confirmed from their spectra. Diamonds indicate galaxies for which IR data is available and well fitted. The dark purple dashed line show the \cite{KennicuttEvans12} relation for normal star-forming galaxies.}
\end{figure}

In Fig.~\ref{lirsfr}, we compare the SFR and L$_{IR}$ of the HRS+COSMOS galaxies obtained by the SED modelling of the UV-IR emission with CIGALE and compare their position with respect to the \cite{KennicuttEvans12} relation.
All the sources are more than a factor of 3 below the relation.
COSMOS sources with a 24\,$\mu$m detection show the same departure from the \cite{KennicuttEvans12} relation with a L$_{IR}$.
The departure from the relation is stronger for the HRS galaxies which are all more than 10 times below it.
For the HRS galaxies, with decreasing timescales of the order of a few hundreds of Myr up to a Gyr, this departure could imply that the IR luminosity is no longer connected to the recent star formation activity and is due to the contribution of the old stellar component.
However, for the COSMOS galaxies and the very short $age_{\mathrm{trunc}}$ values obtained with CIGALE, knowing that CIGALE tends to overestimate them (see Fig.~\ref{mocks}), it is possible that we are probing an evolutionary phase very close in time to the quenching process itself and that the L$_{IR}$ due to the heating by young stars is starting to decrease.
However, for these galaxies too a contribution from the dust heated by evolved stellar populations is also expected and can explain the departure from the \cite{KennicuttEvans12} relation.

\begin{figure}[!h] 
  	\includegraphics[width=\columnwidth]{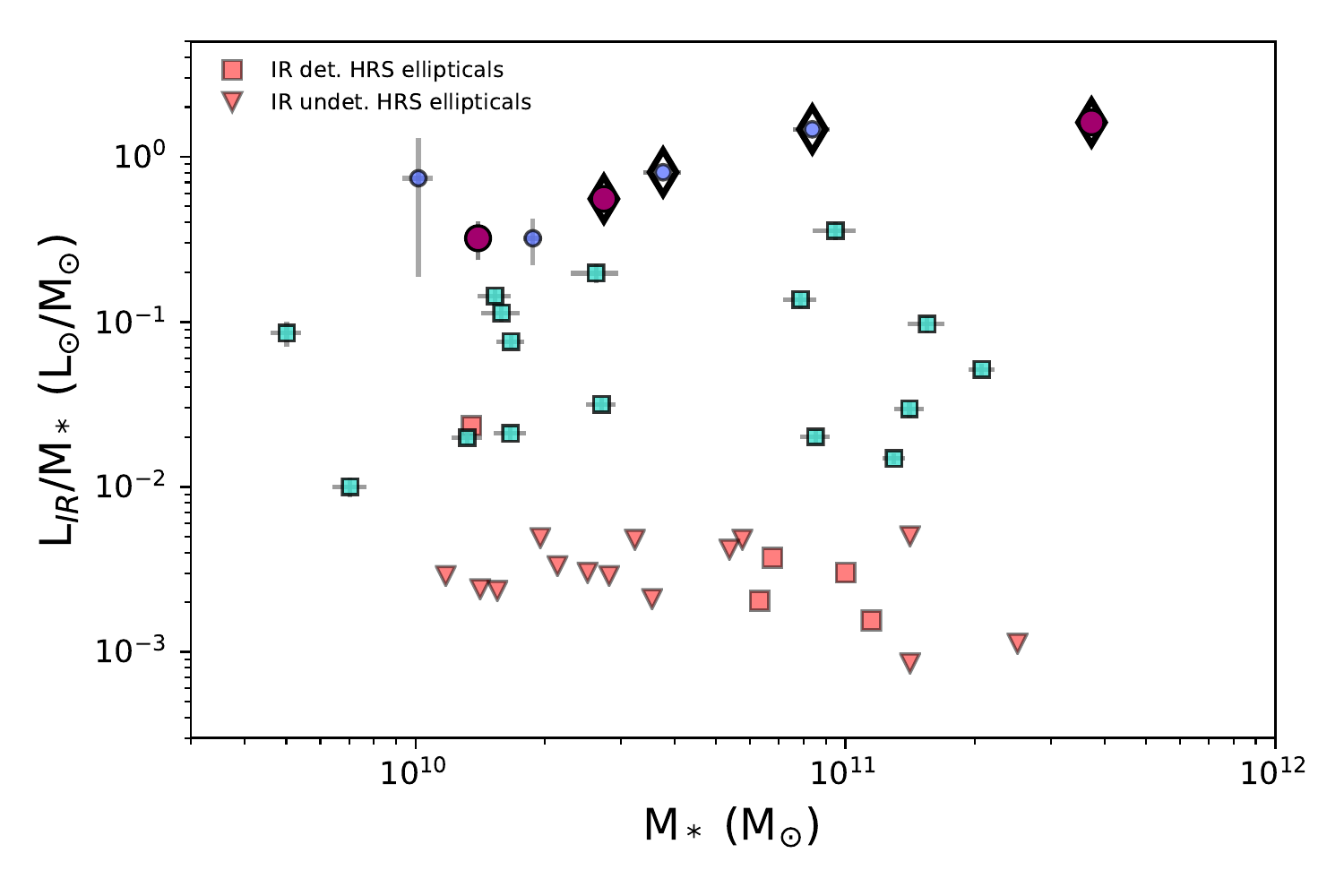}
  	\caption{\label{lirmstar} IR luminosity to stellar mass ratio as a function of stellar mass for the candidate galaxies from both COSMOS and HRS (circles and squares, respectively). In red we show the position of the HRS elliptical galaxies, IR detected (red squares) and IR undetected (red triangles).}
\end{figure}

\begin{figure}[!h] 
  	\includegraphics[width=\columnwidth]{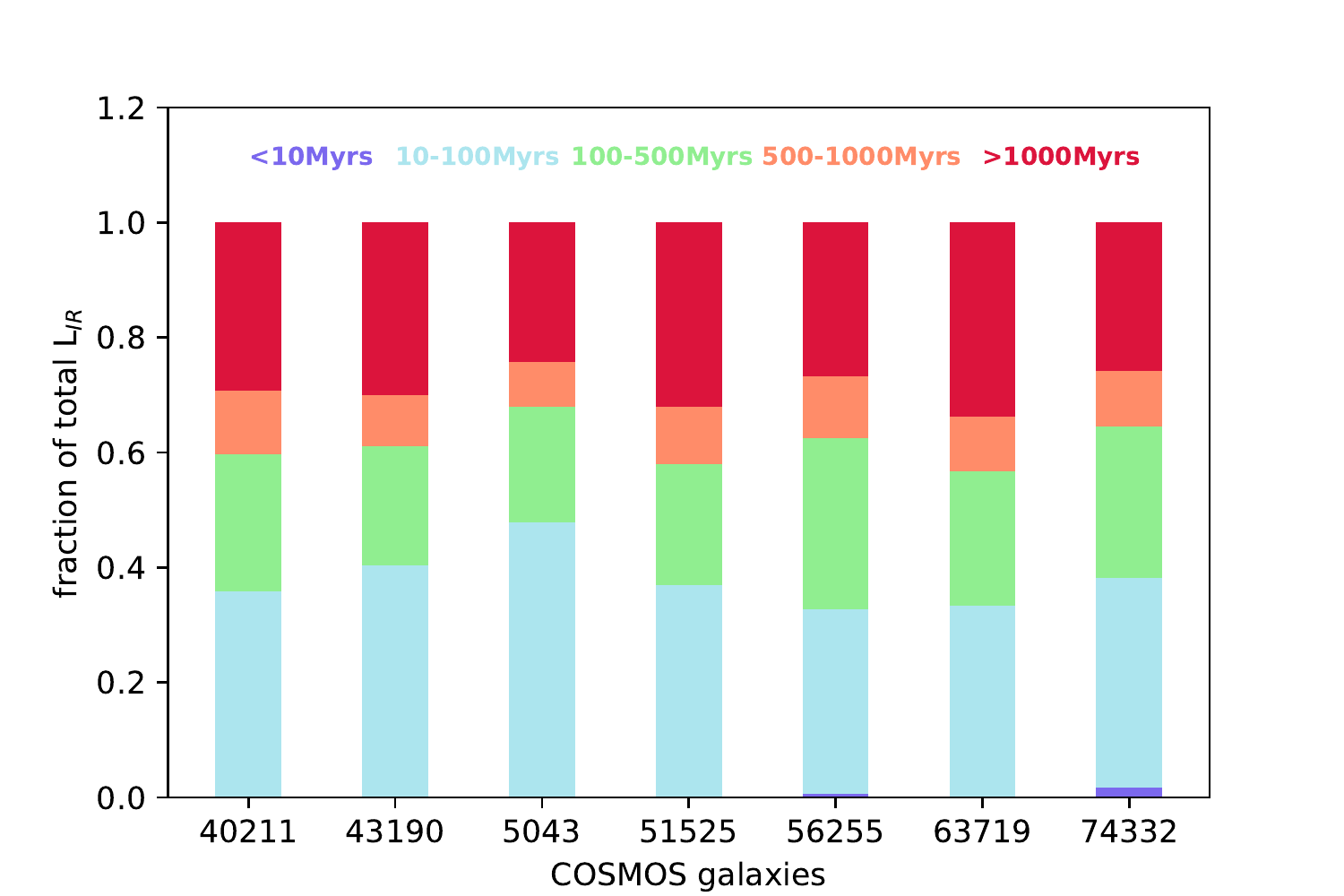}
  	\caption{\label{cont} Contributions of several stellar populations to the total L$_{IR}$ for each of the 7 quenched COSMOS galaxies.}
\end{figure}

To investigate the contribution of the dust heating by evolved stellar population we show in Fig.~\ref{lirmstar} the L$_{IR}/$M$_*$ ratio as a function of stellar mass for our candidates from both HRS and COSMOS.
In addition we show the L$_{IR}/$M$_*$ ratio of HRS elliptical galaxies \citep[see][for the morphology details]{Boselli10b}.
We used their IR data from \cite{Smith12} and \cite{Ciesla12} and the stellar masses from \cite{Boselli10b}.
These sources have no star formation activity and therefore their IR luminosity is due to the dust heating from old stars, or a strong radio AGN \citep{Smith12,Ciesla12,Gomez10,Boselli10a}.
As show in Fig.~\ref{lirmstar} the L$_{IR}/$M$_*$ ratio of our HRS quenched candidates is one to two order of magnitude higher than the typical L$_{IR}/$M$_*$ ratio of elliptical.
For the COSMOS galaxies, the difference is stronger with at least two order of magnitude difference in the L$_{IR}/$M$_*$ ratio between the HRS elliptical and them.
This implies that the IR luminosity that we observe and estimate for our quenched candidates can not be only due to heating from evolved stellar populations.
Although this test rules out the possibility of the L$_{IR}$ coming mainly from dust heated by evolved stellar populations, it does not assess the case of intermediate-age stars which can dominate the dust heating \citep[see e.g.,][]{Utomo14,Hayward14}. 
To estimate the contribution of intermediate age stars in the L$_{IR}$ of COSMOS quenched candidates, we use the SFH obtained from the best fit of each quenched galaxy to quantify the fraction of L$_{IR}$ due to stars in different bins of age using the SED simulation function of CIGALE. 
The results are shown in Fig.~\ref{cont} for different stellar age bins (0-10, 10-100, 100-500, 500-1000\,Myr).
On average, the contribution of 10-100\,Myr stars to the total L$_{IR}$ is 40\% while the contribution of intermediate-age stars (with age between 100 and 500\,Myr) is on average 20-25\%. 
Despite the recent rapid quenching, the fraction of young stars contributing to the L$_{IR}$ is still significant. 

\begin{figure*}[!h] 
  	\includegraphics[width=\textwidth]{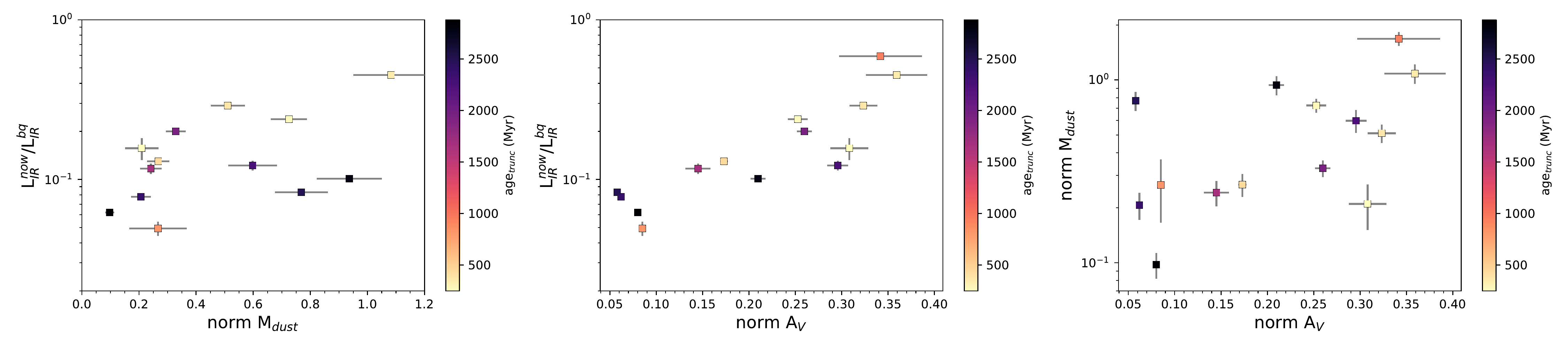}
  	\caption{\label{normparam} Normalised physical properties of the HRS quenched galaxies as a function of L$_{IR}^{now}$/L$_{IR}^{bq}$. They are the dust mass (left panel), the dust temperature (middle panel), and the FUV attenuation (right panel), normalised by the corresponding property of a reference sample built to have the same stellar mass than the quenched source. Symbols are colour-coded by the age of quenching $age_{trunc}$. }
\end{figure*}

The HRS quenched galaxies benefit from a good sampling of their IR SED from MIR to submm (10 flux densities) allowing to measure dust temperatures and dust masses from the SED fitting of the IR range only.
To go a step beyond, we measure those parameters for our quenched candidates. 
Indeed, the effect of ram pressure stripping is the quenching of star formation due to the removal of the gas from the galaxy. 
The HRS galaxies have a quenching age ranging from 100 to 3000\,Myr which is a typical timescale for ram pressure stripping. 
Over several hundreds of Myr, the young stars do no longer heat the dust and therefore the decrease of the L$_{IR}$ is expected. 
There have been observations of dust truncated profiles in galaxies undergoing ram pressure stripping \citep[e.g.,][]{BoselliGavazzi06,Cortese10a,Cortese14,Longobardi20}.
For each of our selected HRS galaxies, we compute the average dust mass of the reference samples of normal star-forming HRS galaxies having the same stellar mass than our selected quenched galaxies.
For each of these sources, we then calculate the ratio between their dust mass and the average one of their reference sample.
In Fig.~\ref{normparam} (left panel), we show the L$_{IR}^{now}$/L$_{IR}^{bq}$ ratio as a function of the normalised dust mass.
A weak trend is seen, but it is clear that they show a deficit in dust content with half of the candidates having a dust mass corresponding to less than 50\% of the dust content of normal galaxies with similar stellar mass.
This result is consistent with the fact that the quenched HRS galaxies have lower attenuation than normal star-forming galaxies of corresponding stellar masses as shown in Fig.~\ref{normparam} (middle panel).
There is a clear trend between L$_{IR}^{now}$/L$_{IR}^{bq}$ and the normalised V band attenuation.
Quenched galaxies with the larger/lower deficit in attenuation compared to their reference sample are those with the lower/larger L$_{IR}^{now}$/L$_{IR}^{bq}$ ratio.
We also computed a ``normalized'' dust temperature for each quenched galaxy using the same reference sample but found no trend.
The normalised dust temperature showed the same value for each quenched galaxies with a large error that can not allow us to conclude on any effect on the heating process.
Either the dynamical range probed by the HRS in terms of dust temperature is not sufficient enough to see any trend, or there is no lower dust temperature in quenched galaxies compared to their reference sample.
Therefore understanding the decrease of L$_{IR}$ is difficult as it can be attributed to both a lack of young stars heating the dust and a deficit in dust content.
Finally, in Fig.~\ref{normparam} (right panel) we show the normalised M$_{dust}$ as a function of the normalised V band attenuation.
Even weak, there is a trend that confirm a link between the lack of dust and the low V band attenuation in the HRS quenched galaxies.
A similar investigation would be needed to understand the origin of the star formation quenching of the COSMOS quenched sources but a better IR sampling of their SED  and/or probes of their gas content  would be needed to perform such an analysis.

\section{\label{conclusions}Conclusions}
We investigate the IR luminosity decrease in galaxies after the quenching of their star formation activity.
First, we use a sample of local well-known galaxies, the \textit{Herschel} Reference Survey by selecting 14 galaxies that experienced a rapid and drastic quenching of their star formation activity (more than 99\%) in the last Gyr.
These galaxies are member of the Virgo cluster and known to have underwent ram pressure stripping.
In addition we selected galaxies at higher (0.5$<z<$1) redshift in the COSMOS field.
We rely on the statistical work of \cite{Aufort20} who provided for a sub-sample of COSMOS galaxies the probability that they underwent a rapid and recent variation of their SFH.
We select 7 sources using the exact same criteria than for the HRS galaxies, that is a decrease of the SFR by more than 99\%.

We perform UV-to-IR SED modelling of the HRS+COSMOS sources to estimate the age of the quenching and the L$_{IR}$ of the galaxies.
For the HRS galaxies, an IR only SED modelling is performed to estimate the present L$_{IR}$ of the galaxies.
We validate our estimate of the age of quenching with the results of \cite{Boselli16} who used spectroscopy in addition of photometry and performed a detailed and more specific analysis of these sources.
For the COSMOS galaxies, we use the measurement of the present L$_{IR}$ provided by the UV-to-IR SED modelling by CIGALE and checked its validity in case of absence of IR data point.

For both the HRS and COSMOS quenched galaxies, we estimate the L$_{IR}$ just before the quenching happened using the SED of each quenched source at the moment right before quenching.
We obtain the past L$_{IR}$ of these galaxies and find that it is consistent with the L$_{IR}$ of a reference sample built for each quenched galaxy to be similar in stellar mass and normally forming stars.
We conclude that our method is able to recover the past properties of galaxies even though the more recent the quenching, the higher the precision on the estimated past L$_{IR}$.

Gathering the local and high-redshift samples, we investigate the relation between the observed L$_{IR}$ to the L$_{IR}$ before quenching ratio and the quenching age to put a constraint on the timescales of the decrease of the IR luminosity after the shutdown of star formation.
Assuming an exponential decrease of the L$_{IR}$ after quenching, we find that the COSMOS quenched galaxies have typical timescales that are short, less than a couple of Myr while the HRS quenched galaxies are consistent with a decrease with an e-folding time of several hundreds of Myr up to 1\,Gyr.
For the HRS quenched galaxies, this is consistent with their known quenching process which is ram pressure stripping due to the environment of the Virgo cluster.
The difference of $age_{\mathrm{trunc}}$ and L$_{IR}$ decreasing timescales between the HRS and the COSMOS quenched galaxies suggest different quenching processes.

A comparison between the L$_{IR}/$M$_*$ ratios of our HRS and COSMOS selected galaxies and those of HRS elliptical galaxies with no more star formation activity show that the IR emission of our candidates can not be only due to dust heating from evolved stellar populations and that the contribution of young stars still contribute to $\sim$40\% of the total L$_{IR}$.
Also, we clearly see a deficit of dust mass in the HRS galaxies, in agreement with a lower attenuation in the V band for the quenched galaxies.  
It is expected as we know that these galaxies suffer from ram pressure stripping that affect the dust content as well.
However, no conclusion can be drawn for the dust temperature with no clear difference seen between the average dust temperature of the quenched candidates and the average one of their reference sample.
In other word, going further in the characterisation of the decrease of IR luminosity is challenging and would need more time sensitive indicators such as IR emission lines.
Further investigations would require probing the dust content of the COSMOS sample to put stronger constraints on the present IR luminosity and probe the dust content of these sources.


\begin{acknowledgements}
L.C. warmly thanks Emanuele Daddi for useful comments that really improved the paper and Mauro Giavalisco for an interesting discussion in Sesto that gave the idea of this study.
The project has received funding from Excellence Initiative of Aix-Marseille
University - AMIDEX, a French ‘Investissements d’Avenir’ programme.
\end{acknowledgements}

\bibliographystyle{aa}
\bibliography{quenching_ir}
\appendix

\section{\label{optspec}Optical spectra of the candidates}
\begin{figure*}[!h] 
   	\includegraphics[width=\textwidth]{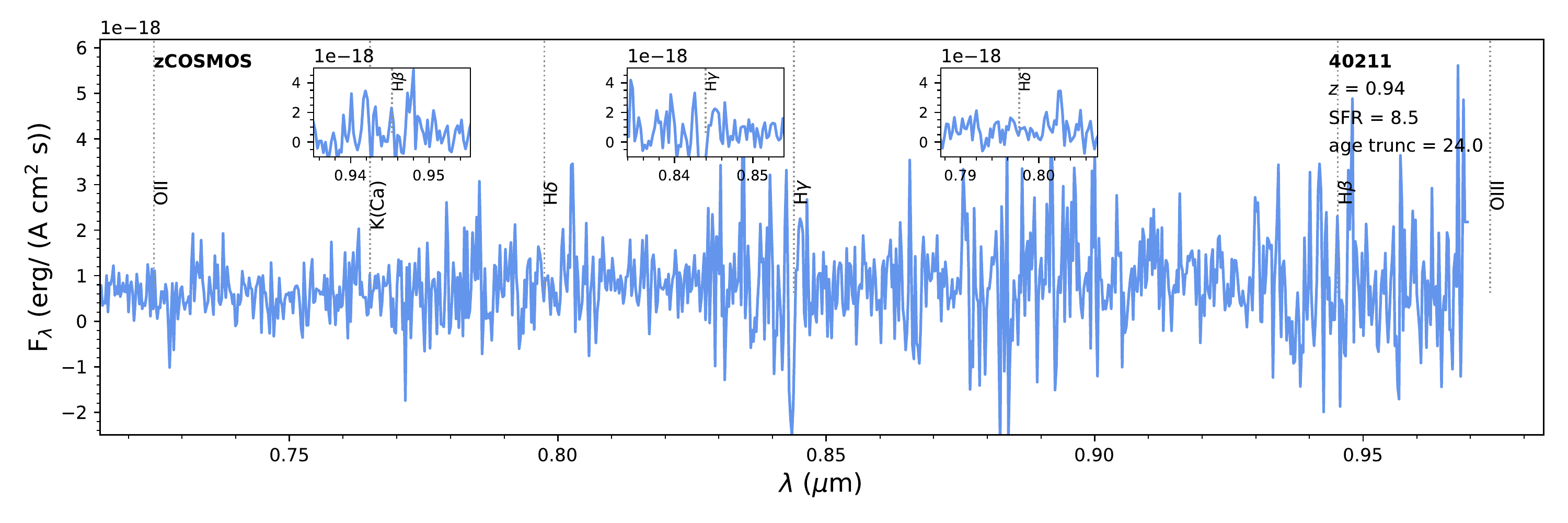}
   	\includegraphics[width=\textwidth]{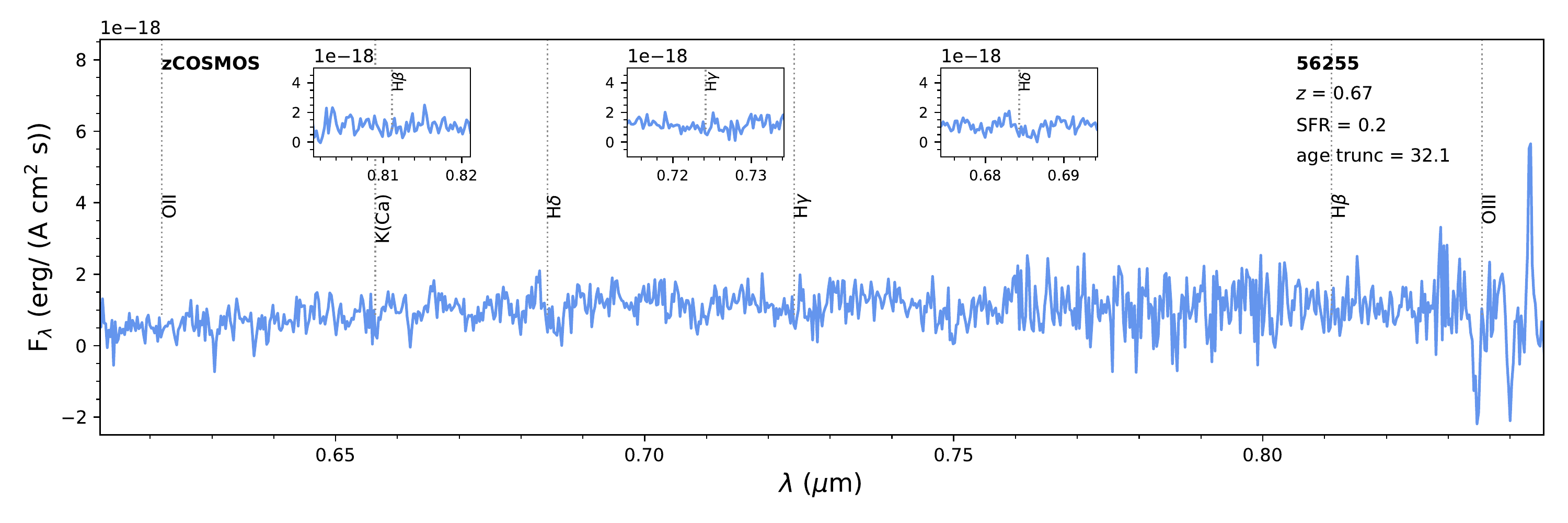}
   	\includegraphics[width=\textwidth]{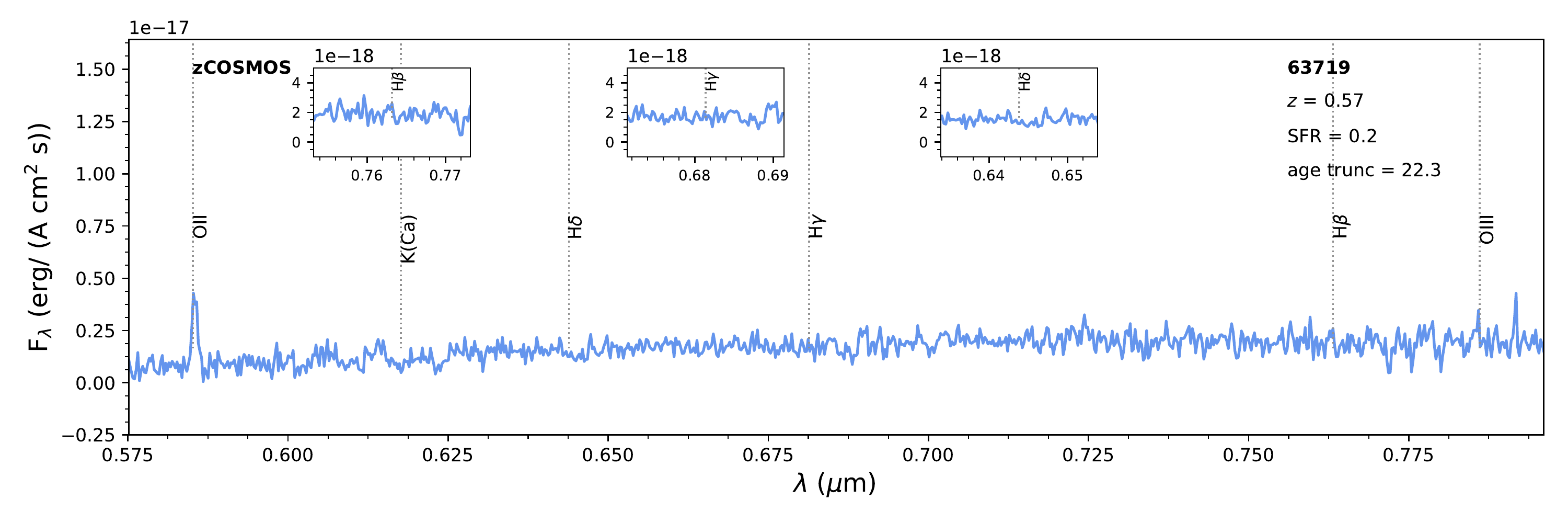}
  	\caption{\label{spec1} zCOSMOS spectra for galaxies of the final sample when available. Inset panels are zoom-in on the H$\beta$, H$\gamma$, and H$\delta$ lines.}
\end{figure*}

\end{document}